\documentclass[preprint]{aastex}

\usepackage{natbib}%,subfig}
\title{Intragroup and Galaxy-Linked Diffuse X-ray Emission in Hickson Compact Groups}

\author{Tyler D. Desjardins}
\affil{Department of Physics and Astronomy, The University of Western Ontario,\\London, ON N6A 3K7, Canada}
\email{tdesjar@uwo.ca}

\author{Sarah C. Gallagher}
\affil{Department of Physics and Astronomy, The University of Western Ontario,\\London, ON N6A 3K7, Canada}

\author{Panayiotis Tzanavaris}
\affil{Laboratory for X-ray Astrophysics, NASA/Goddard Space Flight Center,\\Greenbelt, MD 20771, USA}

\author{John S. Mulchaey}
\affil{Carnegie Observatories, 813 Santa Barbara Street, Pasadena, CA 91101, USA}

\author{William N. Brandt\altaffilmark{1}}
\affil{Department of Astronomy and Astrophysics, The Pennsylvania State University,\\525 Davey Laboratory, University Park, PA 16802, USA}

\author{Jane C. Charlton}
\affil{Department of Astronomy and Astrophysics, The Pennsylvania State University,\\525 Davey Laboratory, University Park, PA 16802, USA}

\author{Gordon P. Garmire}
\affil{Department of Astronomy and Astrophysics, The Pennsylvania State University,\\525 Davey Laboratory, University Park, PA 16802, USA}

\author{Caryl Gronwall\altaffilmark{1}}
\affil{Department of Astronomy and Astrophysics, The Pennsylvania State University,\\525 Davey Laboratory, University Park, PA 16802, USA}

\author{Ann E. Hornschemeier}
\affil{Laboratory for X-ray Astrophysics, NASA/Goddard Space Flight Center,\\Greenbelt, MD 20771, USA}

\author{Kelsey E. Johnson}
\affil{Department of Astronomy, University of Virginia, P.O. Box 3813,\\Charlottesville, VA 22904, USA}

\author{Iraklis S. Konstantopoulos}
\affil{Department of Astronomy and Astrophysics, The Pennsylvania State University,\\525 Davey Laboratory, University Park, PA 16802, USA}

\and

\author{Ann I. Zabludoff}
\affil{Steward Observatory, University of Arizona, 933 North Cherry Avenue,\\Tucson, AZ 95721, USA}

\altaffiltext{1}{Institute for Gravitation and the Cosmos, The Pennsylvania State University, University Park, PA 16802, USA}

\shortauthors{Desjardins et al.}
\shorttitle{Two Sequence Evolution in Hickson Compact Groups}

\newcommand{\degree}{$^{\circ}$}
\newcommand{\e}[2]{$#1\times10^{#2}$}
\newcommand{\hi}{H~{\sc i} }
\newcommand{\rascend}[4]{#1$^{\mathrm{h}}$#2$^{\mathrm{m}}$#3$^{\mathrm{s}}$.#4}
\newcommand{\decline}[4]{#1\degree~#2$\arcmin$~#3~#4$\arcsec$}
\newcommand{\per}[1]{#1$^{-1}$}
\newcommand{\Msun}{M$_\odot$}

\begin{document}

\begin{abstract}
Isolated compact groups of galaxies (CGs) present a range of dynamical states, group velocity dispersions, and galaxy morphologies with which to study galaxy evolution, particularly the properties of gas both within the galaxies and in the intragroup medium. As part of a large, multiwavelength examination of CGs, we present an archival study of diffuse X-ray emission in a subset of nine Hickson compact groups observed with the {\em Chandra X-ray Observatory}. We find that seven of the groups in our sample exhibit detectable diffuse emission. However, unlike large-scale emission in galaxy clusters, the diffuse features in the majority of the detected groups are linked to the individual galaxies, in the form of both plumes and halos likely as a result of star formation or AGN activity, as well as in emission from tidal features. Unlike previous studies from earlier X-ray missions, HCGs~31, 42, 59, and 92 are found to be consistent with the $L_X-T$ relationship from clusters within the errors, while HCGs~16 and 31 are consistent with the cluster $L_X-\sigma$ relation, though this is likely coincidental given that the hot gas in these two systems is largely due to star formation. We find that $L_X$ increases with decreasing group \hi to dynamical-mass ratio with tentative evidence for a dependance in X-ray luminosity on \hi morphology whereby systems with intragroup \hi indicative of strong interactions are considerably more X-ray luminous than passively evolving groups. We also find a gap in the $L_X$ of groups as a function of the total group specific star formation rate. Our findings suggest that the hot gas in these groups is not in hydrostatic equilibrium and these systems are not low-mass analogs of rich groups or clusters, with the possible exception of HCG~62.
\end{abstract}

\section{Introduction}

As the majority of galaxies in the nearby Universe are found within gravitationally bound groups (e.g.,~\citealt{tully87,small99,karachentsev05}), understanding the physical processes at work in these systems is fundamental to galaxy formation, evolution, and cosmological theory. Though a significant fraction may be condensed structures embedded within largely extended, loosely bound systems \citep{tovmassian06,mendel11}, Hickson compact groups (HCGs; \citealt{hickson82}) in particular are useful tools for studying tidally enhanced galaxy evolution in the nearby Universe because they are isolated from other nearby galaxies due to their selection criteria, have low velocity dispersions (usually $\lesssim200$~km~s$^{-1}$), and exhibit high number densities (typically 3--4 galaxies of comparable luminosity within several galaxy radii of one another). These properties combined encourage numerous gravitational interactions between group members. 

Because the crossing time of CGs is short ($\lesssim0.02t_{H_0}$; \citealt{diaferio94}), simulations indicate that group members should have entirely merged into single massive galaxies on timescales of $\sim$1~Gyr. Hypotheses concerning the continued existence of CGs have included that these systems are either recently formed (e.g., \citealt{claudia05}), that they are chance alignments within larger structures (e.g., \citealt{mamon86}), or that CGs are perpetually formed within collapsing rich groups (e.g., \citealt{diaferio94}). However, all three proposed scenarios have evidence that contradicts them such as the wide range in ages of star clusters in CGs \citep{konstantopoulos10,konstantopoulos12}, the low surface brightness tidal debris and disturbed disk morphologies present in many groups (e.g., HCGs 16, 31, and 92; \citealt{claudia98,gallagher10,fedotov11}), and indications that a small fraction of CG members are the result of mergers \citep{claudia05}. Clearly these systems are much more complicated than first suspected, and understanding the origin and physical processes responsible for maintaining CGs over timescales in excess of several Gyr is important for refining cosmological simulations.

\citet{verdes01} presented an evolutionary scenario in which the spiral-rich compact groups with most of their neutral \hi gas confined to the galaxy disks experience successive tidal encounters which liberate the cool gas from the galaxies into the intragroup medium in the form of tails and bridges. This progresses to one of two evolutionary end points: (1) a group in which the liberated \hi is shock-heated to X-ray temperatures; or (2) a group with a large, low velocity dispersion \hi halo. In a study of five groups, \citet{freeland09} found that the amount of \hi tidally removed from galaxy disks is related to the compactness of the group. Therefore, compact groups should be very efficient at dispersing their neutral gas into the intragroup medium. If a sufficient mass of gas is removed from the disks and heated to sufficiently temperatures, the group may be considered a fossil group. \citet{jones03} defined fossil groups as those with $L_X\ge10^{42}~h_{50}^{-2}$~erg~s$^{-1}$ and a difference in $R$-band magnitude of $\Delta m\ge 2.0$~mag between the two brightest group members. The possible evolution of HCGs into fossil groups represents the most similar to clusters these systems may become. We note that tidal stripping of gas from galaxy disks is not necessarily the only source of baryons for the intragroup medium. Simulations of galaxy clusters at high redshift have shown that cold mode accretion of gas from the surrounding intergalactic medium can be a substantial source of baryons (e.g.,~\citealt{keres05}); however, it is unclear how significant this accretion may be for low redshift groups.

Observations of diffuse X-ray emission are particularly helpful in placing constraints on the properties of CGs because the increased likelihood of tidal encounters implies that a significant fraction of the gas mass in these systems may be in the form of a hot plasma. Previous {\em ROSAT} (e.g.,~\citealt{ponman96, mulchaey98, helsdon01, mulchaey03}) and {\em Chandra} (e.g.,~\citealt{rasmussen08, rasmussen09, sun09}) analyses of diffuse X-ray emission in galaxy groups have specifically treated groups of galaxies as low-mass analogs of galaxy clusters. Early studies found that, within large errors, galaxy groups were consistent with the established relationships between bolometric X-ray luminosity $L_X$ and gas temperature ($L_X-T$) as well as $L_X$ and velocity dispersion ($L_X-\sigma$) found from observations of large samples of galaxy clusters. With improved instrumentation and more statistically robust datasets, it was shown that compact groups deviate from the established $L_X-T$ cluster relation (groups are fainter for a given temperature), however the $L_X-\sigma$ relation was still found to be consistent, though with a large scatter (cf.~Figure~12 in \citealt{ponman96}).

Having performed the first extensive X-ray survey of compact groups using data from both the {\em ROSAT} All-Sky Survey (RASS) and pointed {\em ROSAT} PSPC observations, \citet{ponman96} found that 22 out of 85 observed HCGs had diffuse X-ray emission above their detection limits. From the data, the authors statistically estimated that $\sim$75\% of HCGs have a diffuse X-ray luminosity above \e{1.3}{41}~erg~s$^{-1}$. While diffuse X-ray emission was previously thought to be limited to groups containing only E/S0 members, \citet{ponman96} found that groups containing spiral galaxies exhibit diffuse X-ray emission as well, however the X-ray luminosity functions of spiral dominated groups show that they tend to be fainter than E/S0 dominated groups to a high statistical significance. Additionally, a weak anti-correlation was found between $L_X$ and spiral galaxy fraction.

Studies of compact groups such as HCG~31, which shows multiple interactions among its low-mass members, have shown evidence for hierarchical structure formation characteristic of environments at higher redshifts ($z$$\sim$1--2)  \citep{gallagher10}. Additionally, a gap in both the mid-IR colors and specific star formation rates (SSFRs; i.e.,~SFR normalized by stellar mass) of HCGs compared to galaxies in other environments suggests that the galaxies in these systems undergo accelerated evolution from star-forming ``blue-cloud'' galaxies to the quiescent ``red sequence'' \citep{johnson07,walker10,walker12,tzanavaris10}. This rapid evolution is supported by the \hi deficiency observed in most HCG members relative to spiral galaxies in the field \citep{hutchmeier97,verdes01}.

The rapid evolution of HCG galaxies from gas-rich and star-forming to gas-poor and quiescent raises the following question: how is the neutral gas being processed in HCGs, i.e., is most of the \hi mass converted into stellar mass or is it ionized and expelled into the intragroup medium, and in what proportions do these mechanisms operate in individual groups? Previous papers have examined the star cluster populations in several HCGs (e.g.,~\citealt{gallagher01,konstantopoulos10,fedotov11}) to study the consumption of gas by star formation, however there has been no investigation of how the \hi gas is heated to X-ray temperatures in these systems in the context of group evolution. In this work, as part of a multiwavelength campaign to study the evolution of both the galaxies in compact groups and the group environment itself, we present the results of a study of the diffuse X-ray emission from nearby CGs ($z\lesssim0.015$) using {\em Chandra} observations. Specifically, we inspect the morphology of the hot gas in compact groups using improved spatial resolution compared to previous studies that analyzed {\em ROSAT} data; we re-examine the L$_X$ scaling relations with temperature and velocity dispersion; and we compare the hot gas in groups to the neutral \hi gas to investigate how the X-ray gas is built up in these systems.

First, we describe the {\em Chandra} observations and methods for reducing the events files in \S\ref{sec:obs}. In \S\ref{sec:spectra}, we discuss the extraction of the X-ray spectra, as well as the models we used to fit the data. Our model-derived measurements are presented in \S\ref{sec:results} in addition to a discussion of the results of our study in the larger context of galaxy evolution and the evolution of the group environment itself. We summarize our findings and discuss the future steps of our investigation into the diffuse X-ray properties of HCGs in \S\ref{sec:sum}. Appendix~\ref{app:sourcenotes} gives specific information on the extraction and modeling of the diffuse emission in the detected groups. Throughout this work, we assume a cosmology of $H_0=70$~km~s$^{-1}$~Mpc$^{-1}$, $\Omega_{\mathrm{M}}=0.27$, and $\Omega_\Lambda=0.73$. We also caution the reader that we allude several times to compact groups possibly becoming more cluster-like as they evolve. In this context, we refer to the hot gas distribution, not necessarily to the richness of the group; therefore, the term cluster-like is used to describe groups with a large fraction of E/S0 galaxies and a shared, hot intragroup medium.

\section{Sample Definition, Observations, and Data Reduction}
\label{sec:obs}

Objects for study were selected on the basis of data available from the archive at the {\em Chandra} X-ray Center (CXC) for CGs that are part of a sample designed to study star formation and galaxy evolution in the compact group environment (see \citealt{johnson07} for more information regarding the original sample; HCG~92 was subsequently added based on the availability of high-resolution, deep imaging data). To this end, the sample consists of groups at small enough distances ($z\lesssim0.015$, excepting HCG~92) to allow for high spatial resolution imaging over a wide range of wavelengths. In addition to our new {\em Chandra} observations of HCGs~7, 22, 31, and 59 (PIs: Garmire [7, 22] and Gallagher [31, 59]),  we selected archival data that covered the entirety of each group, rather than only individual group members, to search for hot, ionized gas in the intragroup medium, bringing the total number in our study of diffuse X-ray emission to nine groups. HCGs~2, 19, 48, and 61, which were included in the \citet{johnson07} sample, were omitted from our study due to the lack of suitable {\em Chandra} imaging data. 

All of the groups were observed at the ACIS-S aimpoint with the exception of HCG~90, which was observed at the ACIS-I aimpoint due to its large angular extent (the diffuse emission of HCG~90 falls mostly on the I0 CCD, however some is present along the edges of the other I array detectors as well). The individual ACIS CCDs have a field of view of  $8\farcm3\times8\farcm3$, which is comparable to the extent of the diffuse emission in compact groups found in \citet{ponman96} and \citet{mulchaey03} (noting the exceptions of HCGs~62 and 90). All data were taken in VFAINT mode except HCG~16, HCG~62 (obsID 921), and HCG~92 (obsID 789), which were in FAINT mode. Due to the design of the High Resolution Mirror Assembly, observations with the {\em Chandra} ACIS camera have superior spatial resolution ($\sim$0.5\arcsec\ FWHM) compared to other instruments such as the {\em ROSAT} PSPC ($\sim$25\arcsec) and {\em XMM} EPIC ($\sim$6\arcsec), allowing for more robust removal of point sources that contaminate the diffuse emission and can ead to incorrect estimations of hot gas properties. Tables~\ref{tab:targets} and \ref{tab:obs} list the properties and observation information, respectively, of the CGs included our sample.

The data were reprocessed beginning with the level~1 events file. We used the CIAO tool {\tt acis\_reprocess\_events} with standard event filtering and VFAINT background cleaning (when appropriate) to produce the level~2 events file; however, we omitted the pixel randomization step in the event file processing to prevent degradation of the spatial resolution. The pixel randomization introduces a 0\farcs5 random resampling of the event positions on the detector. This procedure is normally performed to mitigate the effects of aliasing in observations $\lesssim2$~ks, however the observations included in our sample are substantially longer than this limit, and therefore the pixel randomization is not required. The pixel randomization has been shown to decrease the spatial resolution of the observations by $\sim$12\% (e.g.,~\citealt{chartas02}). High spatial resolution is useful for proper removal of point sources that may be embedded within or projected onto any diffuse emission present in the groups.

\begin{deluxetable}{lrcrrrrc}
\tablewidth{0pt}
\tabletypesize{\footnotesize}
\tablecolumns{8}
\tablecaption{{\em Chandra} Observation Information\label{tab:obs}}
\tablehead{\colhead{HCG} & \colhead{{\em Chandra}} & \colhead{Date} & \colhead{Extraction Area\tablenotemark{a}} & \colhead{Total Counts\tablenotemark{b}} & \colhead{Exp.} & \colhead{Bkg Counts} & \colhead{Bkg Exp.\tablenotemark{c}}\\
\colhead{} & \colhead{obsID} & \colhead{} & \colhead{(sq. arcmin)} & \colhead{(0.7--3.0 keV)} & \colhead{(ks)} & \colhead{(0.7--3.0 keV)} & \colhead{(ks)}}
\startdata
\object[HCG 7]{7}\tablenotemark{d} & \dataset[ADS/Sa.CXO#obs/08171]{8171}\tablenotemark{d} & 2007-09-13 & 48.2 & 2419 & 19.2 & 57670 & 900 \\
      & \dataset[ADS/Sa.CXO#obs/09588]{9588}\tablenotemark{d} & 2007-09-16 &  & & 16.7 &  &  \\
\object[HCG 16]{16} & \dataset[ADS/Sa.CXO#obs/00923]{923} & 2000-11-16 & 26.9 & 1270 & 12.6 & 6209 & 110 \\
\object[HCG 22]{22}\tablenotemark{e} & \dataset[ADS/Sa.CXO#obs/08172]{8172} & 2006-11-23 & 44.7 & 2078 & 31.8 & 27973 & 450 \\
\object[HCG 31]{31}\tablenotemark{e} & \dataset[ADS/Sa.CXO#obs/09405]{9405} & 2007-11-15 & 8.70 & 636 & 35.6 & 5433 & 450 \\
\object[HCG 42]{42} & \dataset[ADS/Sa.CXO#obs/03215]{3215} & 2002-03-26 &  36.5 & 7600 & 31.7 & 15882 & 450 \\
\object[HCG 59]{59}\tablenotemark{e} & \dataset[ADS/Sa.CXO#obs/09406]{9406} & 2008-04-12 & 24.7 & 1457 & 38.4 & 15373 & 450 \\
\object[HCG 62]{62} & \dataset[ADS/Sa.CXO#obs/00921]{921} & 2000-01-25 & 7.54 & 31053 & 48.5 & 2604 & 145 \\
   & \dataset[ADS/Sa.CXO#obs/10462]{10462}\tablenotemark{d} & 2009-03-02 & 7.54 & 55302 & 67.1 & 247702 & 900 \\
   & \dataset[ADS/Sa.CXO#obs/10874]{10874}\tablenotemark{d} & 2009-03-02 &  &  & 51.4 &  &  \\
\object[HCG 90]{90} & \dataset[ADS/Sa.CXO#obs/00905]{905} & 2000-07-02 & 68.3 & 24821 & 49.5 & 166640 & 450 \\
\object[HCG 92]{92} & \dataset[ADS/Sa.CXO#obs/00789]{789} & 2000-07-09 & 19.3 & 2545 & 19.7 & 4663 & 110 \\
  &  \dataset[ADS/Sa.CXO#obs/07924]{7924} & 2007-08-17 & 19.3 & 10656 & 93.2 & 27337 & 450 \\
\enddata
\tablenotetext{\ }{{\bf Notes.}}
\tablenotetext{a}{Total area over which the events were extracted}
\tablenotetext{b}{Source + Background in science observation}
\tablenotetext{c}{Exposure time of the ACIS stowed background files}
\tablenotetext{d}{Count information listed for merged spectra}
\tablenotetext{e}{Counts evaluated over 0.7--1.7 keV (see \S\ref{sec:extract})}
\end{deluxetable}

Because the diffuse intragroup emission is typically weak and there is no robust method to determine the full radial extent of it from the data, for most targets we did not perform a spectral extraction in an annulus surrounding the targets to determine the soft X-ray background level. Instead, we subtracted the instrumental background using the stowed ACIS background files appropriate to each observation (as in \citealt{white03}). The stowed ACIS background files\footnote{This brief description of the stowed ACIS background is based on the work of Maxim Markevitch; further details are at \url{http://cxc.harvard.edu/contrib/maxim/stowed/}.} were obtained by sliding ACIS out of view of the sky and away from the external calibration source to avoid spectral-line contamination. All stowed background observations were taken after 2002 in VFAINT mode with gain corrections applied. Following procedures outlined in the ACIS Extract manual \citep{broos10}, we removed the ``Clean55'' bit from the event lists for subtraction from data taken in FAINT mode and renormalized the background files to the data using the particle background in the range 9--12~keV. The stowed background files include emission from the high energy particle background ($E\gtrsim2$~keV), but not the soft X-ray background (SXRB; $E\lesssim2$~keV), therefore in groups where we used the stowed background files we expect contamination from the SXRB at energies $<$2~keV. Table~\ref{tab:obs} lists total counts extracted from and exposure times of the stowed background data for each group.

We used the latest version of ACIS Extract to create exposure maps, model point source PSFs and excise them from the observations, and extract spectral information (see \S{\ref{sec:spectra}}). The energy used to create the exposure maps was the median event energy over the range 0.5--3.0~keV, chosen to coincide with the peak of any soft diffuse emission in the data.

\section{Spectral Extraction and Modeling}
\label{sec:spectra}

\subsection{Point Source Detection and Removal}

We used the CIAO Mexican-hat wavelet detection routine {\tt wavdetect} \citep{freeman02} to search for point sources in the field of view down to a source-significance threshold of $10^{-5}$. Detections were examined by eye to remove spurious sources (e.g., those with an axis length $\lesssim1$~pixel). We then used ACIS Extract to model the PSF of the sources with MARX prior to excising the point sources from the events files. The excised regions correspond to 1.1 times the 99\% encircled energy radius. For extraction of spectra, we did not interpolate over the holes created by excising the point sources to avoid making statistical assumptions concerning the gas. The area of diffuse emission on the sky lost due to point-source removal was typically less than 1\% of the total extraction region. We defer the examination of the point sources in our sample to Tzanavaris~et~al.~({\em in prep}).

\subsection{Extraction of the Diffuse Emission}
\label{sec:extract}

Extraction regions were centered on the apparent centers of the CG galaxy distributions with shapes and sizes chosen to best cover all of the main group members on the S3 chip (I0 for HCG~90), including evident diffuse emission in the level~2 events files. The extraction regions were either circular or elliptical in shape, except in the case of HCG~16, which fills approximately half of the S3 CCD. In this group, the region was rectangular and placed at an angle to include the main galaxies and the known distribution of \hi gas. The area of each extraction region can be found in Table~\ref{tab:obs}, while additional information is located in Appendix~\ref{app:sourcenotes}. Note that in the case of HCG~22, the extraction region includes a background pair of galaxies to the southeast of the group center. As this group was classified as a non-detection, we set an upper limit on the diffuse X-ray luminosity using the method outlined below. Therefore, the inclusion of this pair in the extraction region has negligible impact on the results.

Prior to model fitting in XSPEC \citep{arnaud96}, the extracted spectra were rebinned such that each pulse height amplitude (PHA) bin contained a minimum of 20 counts. This ensured statistically valid results when fitting with $\chi^2$ statistics. In cases where the ACIS instrumental response did not change significantly between observations (i.e., multiple obsIDs within an observing cycle), we merged the on-source spectra to reduce the relative error in each PHA bin. When merging the spectra was not possible due to the changes in the instrument response, we simultaneously fit the model to each spectrum with the temperature and metal abundances linked between different obsIDs; however, the model normalization was allowed to vary freely.

A detection was defined to have $\mathrm{S/N}\ge3$. The noise of the spectra was determined to be \linebreak$\sigma=\left[SB+(A_st_s/A_bt_b)B\right]^{1/2}$, where $SB$ is the total counts in the source before background subtraction, $B$ is the number of counts in the background, $A$ is the area of the extraction region, $t$ is the integration time, and the subscripts $s$ and $b$ represent the science and background observations, respectively. When using the stowed background, the background counts were scaled to the 9--12~keV count rates in the science observations prior to estimating the noise. This scaling has the effect of normalizing the shape of the stowed background spectrum to that of the source in a range of energies dominated by the particle background \citet{hickox06}. When no diffuse emission was detected, we put an upper limit on the luminosity of the intragroup medium by fixing the temperature and metal abundance of the plasma to reasonable values of 0.6~keV and 0.5~Z$_\odot$, respectively. We then adjusted the normalization to match the sum of the observed count rate and the 1$\sigma$ noise estimate. 

In all cases when diffuse X-ray emission was not detected, the energy range over which the count rates were evaluated had to be restricted to 0.7--1.7~keV due to oversubtraction of the stowed background between 1.7 and 2~keV. Bolometric X-ray luminosities were then computed in the same manner as for groups with detected emission (see below).

\subsection{Spectral Model Fitting}
\label{subsec:modeldescr}

We fit a combination of optically thin plasma and Galactic photoelectric absorption models for each of the groups. The best fitting models are presented in Table~\ref{tab:models}. The first model component, an optically thin plasma, was modeled using the MEKAL model \citep{mewe85,mewe86,kaastra92,kaastra93,liedahl95,kaastra95} with the adopted ionization balance taken from \citet{arnaud85} and \citet{arnaud92}. We fixed the hydrogen number density at a reasonable value of 1~cm$^{-3}$ and allowed the remaining parameters to vary freely. While the metallicity parameter was allowed to vary, these values are poorly constrained by the available data, and we only report them here to describe the best-fitting models to the observations. We calculated the model at all temperatures rather than interpolating it from a pre-calculated table while fitting the data. Other than the plasma temperature, we also report the normalization used in calculating the X-ray luminosity. We note that because the gas is often associated with individual galaxies rather than a single distribution permeating the intragroup medium (see \S\ref{sec:morph}), it is likely multi-temperature; however, we find that the single temperature plasma model fits the extracted spectra well, except in the case of HCG~62 for which a two-temperature plasma results in a better fit. Note that HCG~62 has the most counts in the background-subtracted data compared to any other observation in this study; specifically, a factor of $\sim$7 more counts in the merged obsIDs 10462 and 10874 compared to HCG~90, which has the next most counts. For other targets, our ability to isolate the multiple spectral components that likely make up the diffuse X-ray emission is limited by the number of counts; therefore, it is probable that we cannot detect spectral complexity in our data (e.g.,~multi-temperature plasmas) given the limited numbers of counts.

\begin{deluxetable}{lcrrrrr}
\tablewidth{0pt}
\tabletypesize{\small}
\tablecolumns{7}
\tablecaption{Best Fit Spectral Model Parameters\label{tab:models}}
\tablehead{\colhead{HCG\tablenotemark{a}} & \colhead{\hi} & \colhead{kT} &%
\colhead{$Z/Z_\odot$} & \colhead{$A_{MEKAL}$} & \colhead{log$_{10}(L_X)$} & \colhead{$\chi^2/$D.O.F.}\\
\colhead{} & \colhead{($10^{20}$ cm$^{-2}$)} & \colhead{(keV)} & \colhead{} & %
\colhead{(10$^{-4}$ cm$^{-5}$)} & \colhead{(erg s$^{-1}$)} & \colhead{}}
\startdata
7\tablenotemark{b} & 1.97 & 0.6 & 0.5 & $<0.29$ & $<40.35$ & $\cdots$ \\
16 & 2.56 & $0.65\pm0.06$ & 0.16 & $3.90^{+2.20}_{-1.96}$ & $41.10^{+0.19}_{-0.30}$ & 40.43/47 \\
22\tablenotemark{b} & 4.26 & 0.6 & 0.5 & $<0.21$ & $<39.84$ & $\cdots$ \\
31 & 5.70 & $0.65^{+0.18}_{-0.31}$ & 0.05 & $1.16^{+3.48}_{-0.65}$ & $40.54^{+0.60}_{-0.36}$ & 31.43/40 \\
42 & 4.11 & $0.72\pm0.01$ & 0.55 & $8.38^{+1.76}_{-1.75}$ & $41.82^{+0.08}_{-0.10}$ & 196.48/114 \\
59 & 2.64 & $0.29^{+0.41}_{-0.05}$ & 17.53 & $0.01\pm0.004$ & $40.08^{+0.12}_{-0.16}$ & 55.84/61 \\

62 (921) cool & 3.32 & $0.71\pm0.01$ & 16.68 & $0.50^{+1.23}_{-0.49}$ & $41.88^{+0.54}_{-1.81}$ & 319.10/281 \\
62 (921) hot & 3.32 & $1.20\pm0.04$ & 0.48 & $1.46^{+1.47}_{-1.16}$ & $42.00\pm0.04$ & 319.10/281 \\
62 (921) total &  &  &  &  & $42.23^{+0.34}_{-0.26}$ &  \\
 
62 (10462+10874) cool & 3.32 & $0.71\pm0.01$ & 16.68 & $0.46^{+1.12}_{-0.45}$ & $41.84^{+0.54}_{-1.81}$ & 319.10/281 \\
62 (10462+10874) hot & 3.32 & $1.20\pm0.04$ & 0.48 & $1.48^{+1.43}_{-1.08}$ & $42.01^{+0.04}_{-0.03}$ & 319.10/281 \\
62 (10462+10874) total &  &  &  &  & $42.23^{+0.31}_{-0.25}$ &  \\
 
62 average total &  &  &  &  & $42.72^{+0.33}_{-0.26}$ &  \\
90 BCD & 2.02 & $0.66\pm0.03$ & 0.54 & $2.66_{-1.50}^{+1.59}$ & $40.79^{+0.20}_{-0.36}$ & 121.75/119 \\
92 (789) & 6.16 &  $0.63\pm0.02$ & 0.14 & $6.25_{-0.99}^{+1.08}$ & $41.77^{+0.07}_{-0.08}$ & 357.33/220 \\
92 (7924) & 6.16 & $0.63\pm0.02$ & 0.14 & $5.81^{+1.28}_{-1.12}$ & $41.73^{+0.09}_{-0.10}$ & 357.33/220 \\
92 average &  &  &  &  & 41.76$\pm$0.07 & 
\enddata
\tablenotetext{\ }{\bf Notes.}
\tablenotetext{a}{For groups with spectra that were fitted simultaneously rather than merged, we list the corresponding obsIDs in parentheses}
\tablenotetext{b}{No detection, model temperature and abundance fixed at 0.6~keV and 0.5~Z$_\odot$, respectively}
\end{deluxetable}

Photoelectric absorption was modeled using the Tuebingen-Boulder ISM absorption model from \citet{wilm00}. The only model parameter, the \hi column density along the line of sight, was fixed at the value determined using the HEASoft tool {\tt nH} to compute the weighted mean of \hi in a cone centered on the source and with a radius of 1\arcdeg. Following the recommendation of the {\tt nH} manual\footnote{\url{http://heasarc.nasa.gov/Tools/w3nh_help.html\#comparison}}, we use the \hi values from \citet{kalberla05}.

Abundances and depletion values for the model (used in both the absorption and emission components) were taken from \citet{lodders03} (the most recent abundances available in XSPEC) rather than the default from \citet{anders89}. Both abundance tables use Solar photospheric line and CI chondrite analyses to determine the relative amounts of each element. We found that the values from \citet{lodders03} consistently performed better at fitting emission-line features in the spectra, particularly in observations with significant numbers of counts (e.g.,~HCG~62).

Because {\em Chandra} only has significant response over the energy range 0.3--8~keV, we computed the bolometric X-ray luminosity $L_X$ using a dummy response over the energy range 0.01--100~keV logarithmically divided into 5,000 energy bins. We used 3K~CMB adjusted velocities \citep{fixsen96} from the NASA Extragalactic Database (NED), which adjusts the velocities for the observed dipole anisotropy in the CMB, to determine the distances to the sources for luminosity calculations. Parameters of interest, specifically the temperature and luminosity of the plasma (and by necessity, the model normalization), are reported with 90\% confidence error bars. For simultaneously fit spectra, we report the average luminosity (weighted by the number of counts) determined from the model fits. Note that in the case of HCG~59, the peak in energy of the X-ray emission is very poorly constrained; therefore we fix the temperature to the best-fitting value prior to determining the error in the model normalization. Furthermore, we do not consider the best-fitting value to be representative of the real temperature and simply report it as $<1$~keV, however we do use the temperature result of the model fit for qualitative purposes in the figures below. We present the temperatures and luminosities derived from the fitted models, as well as goodness of fit estimations, in Table~\ref{tab:models}.

\begin{deluxetable}{lrrrrrrrrrrrr}
\rotate
\tablecolumns{13}
\tabletypesize{\footnotesize}
\tablewidth{0 pt}
\tablecaption{Properties of Sample HCGs\label{tab:targets}}
\tablehead{\colhead{HCG} & \colhead{Redshift} & \colhead{$v_{\mathrm{CMB}}$\tablenotemark{a}} & \colhead{Dispersion} & \colhead{Velocity} & \multicolumn{3}{c}{Number of Galaxies} & \colhead{SSFR\tablenotemark{c}} & \colhead{log$_{10}$(M$_{\mathrm{dyn}}$)} & \colhead{log$_{10}$(M$_{\mathrm{H~{I}}})$} & \colhead{log$_{10}$(M$_{\mathrm{H~{I}}})$/} & \colhead{Evo}\\
\colhead{} & \colhead{} & \colhead{(km~s$^{-1}$)} & \colhead{(km s$^{-1}$)} & \colhead{References} & \colhead{Main} & \colhead{Dyn\tablenotemark{b}} & \colhead{E/S0} & \colhead{(10$^{-11}$~yr$^{-1}$)} & \colhead{(M$_{\odot}$)} & \colhead{(M$_{\odot}$)} & \colhead{log$_{10}$(M$_{\mathrm{dyn}}$)\tablenotemark{d}} & \colhead{Type\tablenotemark{e}}}
\startdata
7   & 0.0141 & 3885 & 129$^{+9}_{-8}$ & 1--4  & 4 & 5 & 1 & 8.79$\pm$0.75 & 12.10 & 9.76 & 0.81 & II A  \\
16 & 0.0132 & 3706 & 84$^{+5}_{-4}$ & 4--7 & 5 & 7 & 0 & 56.19$\pm$7.13 & 11.73 & $>10.42$ & $>0.89$ & II B   \\
22 & 0.0090 & 2522 & 37$^{+13}_{-8}$ & 8--11 & 3 & 4 & 1 & 6.24$\pm$0.95 & 10.78 & 9.13 & 0.85 & II A  \\
31 & 0.0135 & 4026 & 56$^{+5}_{-7}$ & 1, 12--14 & 4 & 8 & 0 & 89.63$\pm$15.46 & 10.64 & 10.27 & 0.97 & I B   \\
42 & 0.0133 & 4332 & 273$\pm$12 & 15, 16 & 4 & 38 & 4 & 0.74$\pm$0.11 & 12.75 & 9.40 & 0.74 & III B \\
59\tablenotemark{f} & 0.0135 & 4392 & 208$\pm$12 & 17--20 & 4 & 8 & 1 & 51.82$\pm$11.04 & 12.19 & 9.49 & 0.78 & III \\
62\tablenotemark{f} & 0.0137 & 4443 & 398$\pm$7 & 7, 11, 15, & 4 & 62 & 4 & 0.92$\pm$0.19 & 12.86 & $<9.06$ & $<0.70$ & III \\
     &               &            &                       & 16, 21--23           &             &     &                             &            &                 &                 & \\
90 & 0.0088 & 2364 & 177$^{+8}_{-9}$ & 8, 15 & 4 & 16 & 2 & $\cdots$ & 12.19 & 8.70 & 0.71 & III A \\
92 & 0.0215 & 6119 & 343$\pm$6 & 1, 12, 20 & 4 & 4 & 2 & $\cdots$ & 12.75 & 10.23 & 0.80 & II B 
\enddata
\tablenotetext{\ }{{\bf Notes.}}
\tablenotetext{a}{Velocity measured relative to the 3K CMB}
\tablenotetext{b}{Number of galaxies used in velocity dispersion calculation}
\tablenotetext{c}{Total UV+24~$\mu$m specific star formation rates by \citet{tzanavaris10} with corrections from Tzanavaris~(2012,~private communication)}
\tablenotetext{d}{Dynamical masses estimated from the velocity dispersions listed in column 3 using the median two galaxy separation for the group radius \citep{hickson92} corrected to our cosmology; The most recent \hi masses were taken from \citet{verdes01} and \citet{borthakur10}}
\tablenotetext{e}{Defined by \citet{johnson07} using the ratio of the group \hi mass to its dynamical mass, with subtypes A and B qualitatively assessing the location of the \hi gas (localized to the group members or spread throughout the group, respectively) \citep{konstantopoulos10}}
\tablenotetext{f}{No \hi imaging data available to estimate subtypes in the evolutionary classification}
\tablenotetext{\ }{{\bf References.}}
\tablenotetext{\ }{(1)~\citet{nishiura00}; (2)~\citet{konstantopoulos10}; (3)~SDSS Early Release; (4)~SDSS~DR1; (5)~\citet{ribeiro96}; (6)~\citet{paturel03}; (7)~\citet{theureau98}; (8)~\citet{decarvalho97}; (9)~\citet{huchtmeier94}; (10)~\citet{huchra93}; (11)~\citet{monnier03}; (12)~\citet{rc3}; (13)~\citet{mendes06}; (14)~\citet{verdes05}; (15)~\citet{zabludoff98}; (16)~\citet{zabludoff00}; (17)~\citet{falco99}; (18)~SDSS~DR4; (19)~SDSS~DR8; (20)~\citet{hickson92}; (21)~\citet{dacosta98}; (22)~HIPASS Final Catalog; (23)~\citet{jones09}}
\end{deluxetable}

\clearpage
\begin{deluxetable}{lrrrrrr}
\tablecolumns{7}
\tablewidth{0pt}
\tabletypesize{\footnotesize}
\tablecaption{Comparison of {\em Chandra} and {\em ROSAT} Measurements \label{tab:ponmancompare}}
\tablehead{\colhead{HCG} & \colhead{$T_{X, \mathrm{P96}}$} & \colhead{$T_X$} & \colhead{$\Delta T_X$} & \colhead{$F_{X,{\mathrm P96}}$\tablenotemark{a}} & \colhead{$F_X$\tablenotemark{a}} & \colhead{$\Delta F_X$}\\
 & \colhead{(keV)} & \colhead{(keV)} & \colhead{(keV)} & \colhead{(10$^{-12}$ erg s$^{-1}$ cm$^{-2}$)} & \colhead{(10$^{-12}$ erg s$^{-1}$ cm$^{-2}$)} & \colhead{(10$^{-12}$ erg s$^{-1}$ cm$^{-2}$)}}
\startdata
16 & $ 0.30 \pm 0.05 $ & $ 0.65 \pm 0.06 $ & $ -0.35 \pm 0.08 $ & $ 7.97^{+1.18}_{-1.03} $ & $ 1.98^{+1.34}_{-0.99} $ & $ 5.99^{+1.78}_{-1.43} $ \\                                                       
42 & $ 0.82 \pm 0.03 $ & $ 0.72 \pm 0.01 $ & $ 0.10 \pm 0.03 $ & $ 23.71^{+1.12}_{-1.07} $ & $ 20.74^{+4.38}_{-4.34} $ & $ 2.97^{+4.52}_{-4.47} $ \\                                                      
62\tablenotemark{b} & $ 0.96 \pm 0.04 $ & $ 0.99 \pm 0.04 $ & $ -0.03 \pm 0.06 $ & $ 169.48^{+12.12}_{-11.31} $ & $ 158.30^{+182.19}_{-70.78} $ & $ 11.18^{+182.59}_{-71.68} $\\                                          
90 & $ 0.68 \pm 0.12 $ & $ 0.66 \pm 0.03 $ & $ 0.02 \pm 0.12 $ & $ 11.34^{+2.61}_{-2.12} $ & $ 4.47^{+2.67}_{-2.53} $ & $ 6.87^{+3.73}_{-3.30} $\\                                                       
92 & $ 0.75 \pm 0.08 $ & $ 0.63 \pm 0.02 $ & $ 0.12 \pm 0.08 $ & $ 9.04^{+0.87}_{-0.80} $ & $ 6.95^{+1.20}_{-1.09} $ & $ 2.09^{+1.48}_{-1.35} $\\                                                        
\enddata
\tablenotetext{\ }{{\bf Notes.}}
\tablenotetext{\ }{P96 = \citet{ponman96}}
\tablenotetext{a}{Converted from the bolometric luminosity and adjusted for differences in cosmology}
\tablenotetext{b}{The values listed for this study are the flux-weighted average temperature and total flux of the two plasma components in HCG~62}
\end{deluxetable}

\clearpage
\section{Results and Discussion}
\label{sec:results}

We identify hot diffuse gas in seven of nine groups (HCGs 16, 31, 42, 59, 62, 90, and 92) in our X-ray sample.  Of the detected groups, diffuse emission in HCG~31 has not been reported in previous studies. Both of the two groups without detections, HCGs~7\footnote{A deeper 49~ks {\em XMM} observation (PI: Belsole) of this group exists, but no results have been published at the time of this writing. A brief inspection of the data shows several dozen point sources and large-scale diffuse emission across the field of view of the EPIC camera ($\sim$5 times the angular size of HCG~7). The sensitivity of {\em XMM} to cool, extended gas complicates the interpretation of this emission as either related to the group or part of the SXRB. Results from this and other {\em XMM} data will be included in subsequent papers.} and 22, have low to negligible star formation without much evidence of strong tidal interactions in the past few Gyr (e.g., \citealt{konstantopoulos10}). In the detected groups, the temperatures are all fairly similar (0.6--0.72~keV not including the hot component of HCG~62 or the anomalously low temperature of HCG~59, which is very poorly constrained), while the range in X-ray luminosities spans $10^{40.37}$--$10^{42.18}$~erg~s$^{-1}$. The morphology of the hot gas ranges from isolated around individual group members to common X-ray halos (e.g., HCGs 16 and 62, respectively), as well as gas bridges connecting galaxies (HCGs 59 and 90; see \S\ref{sec:morph}). In Figures~\ref{fig:maps}--\ref{fig:ssfr_lx}, we plot all values derived from model fits for multiple observations that could not be merged prior to fitting.

Table~\ref{tab:targets} lists relevant information about each of the groups in our sample. Included in the table are the group redshift and 3K~CMB velocity, the calculated velocity dispersion, number of main group members, the number of galaxies used in the velocity dispersion calculation, the number of E/S0 type galaxies that are not considered dwarf members, the specific star formation rate from \citet{tzanavaris10}, the total \hi mass from \citet{verdes01} and \citet{borthakur10}, the dynamical mass determined from the group velocity dispersion and mean two galaxy separation taken from \citet{hickson92}, and the group \hi evolutionary type defined by \citet{johnson07} (see \S\ref{subsec:hi} for a full description). We will continually refer back to these data throughout the figures and discussion that follow.

The nine groups in our sample were also observed with {\em ROSAT} and the results presented in \citet{ponman96}. We compare our temperatures and luminosities derived from fitting the extracted spectra from the {\em Chandra} data with those also detected by \citet{ponman96} (i.e., HCGs~16, 42, 62, 90 and 92) in Table~\ref{tab:ponmancompare}. Rather than directly compare the luminosities, we calculate the fluxes of the sources, which removes the dependence of the assumed cosmology. The redshifts used in the flux calculation for the {\em ROSAT} data are the same as those listed in the redshift column of Table~\ref{tab:targets}. Note that \citet{ponman96} give errors in the luminosity and temperature at the 1$\sigma$ level. For the temperature in HCG~62, we use the luminosity-weighted average of the two plasma temperatures, while the fluxes in HCGs~62 and 92 are based on the average of the luminosities weighted by total counts in the different {\em Chandra} obsIDs, therefore the percent errors on these values seem quite large because of the compounding of errors from multiple measurements. In all cases, the luminosity of the X-ray emission from the {\em ROSAT} data is found to be brighter in the \citet{ponman96} study. When comparing the temperatures we determined against those from \citet{ponman96}, we find that HCGs~62 and 90 are consistent between the two studies, while HCG~16 is significantly hotter, and HCGs~42 and 92 are significantly cooler than previously reported. The mean temperature of the groups detected in both studies is 0.70 and 0.72~keV with standard deviations of 0.25 and 0.14~keV in \citet{ponman96} and this study, respectively. The temperature discrepancies are likely due to the difference in the PSF of {\em ROSAT} PSPC instrument compared to the {\em Chandra} ACIS camera. Specifically, the detection and subtraction of point sources in the diffuse emission is much more robust using {\em Chandra} data, and point sources contaminating the extracted spectrum would alter the peak of the emission leading to an incorrect estimation of the hot gas temperature. We also note that \citet{ponman96} attempted to subtract the diffuse X-ray contribution of the individual group members and interpolate over the `holes'. This, combined with distinctions in the extraction regions and the responses of the {\em ROSAT} PSPC and {\em Chandra} ACIS instruments, may explain the differences in the results.

\subsection{Gas Distribution and Morphology}
\label{sec:morph}

Figure~\ref{fig:maps} shows the smoothed diffuse X-ray emission in the CGs included in our study. To construct the X-ray maps, we smoothed the level~2 events files in the range of 0.5--2~keV using the adaptive smoothing {\tt csmooth} algorithm. Pixels with S/N $>3$ above the stowed background were smoothed with a Gaussian kernel with smoothing scales between 2 and 10 pixels. From the figures, we note that the hot gas in compact groups exhibits varied morphologies including small halos around one or several galaxies, plumes centered on particular group members, tidal bridges, and large common halos encompassing most of the group. 

The maps were qualitatively assessed by eye to categorize the observed X-ray emission as either associated with the environment or the individual group members. Of the nine CGs included in our study, we do not detect diffuse emission from HCGs~7 and 22. From those groups detected with {\em Chandra}, only HCG~62 has emission that permeates the IGM, while HCG~42 has a bright X-ray halo centered on the brightest group galaxy (i.e.,~42A). We consider these two groups indicative of systems with X-ray emission similar in morphology to clusters. The other groups, HCGs~16, 31, 59, 90, and 92 (excepting the shock front in the case of HCG~92), are those in which the hot gas is associated with the individual galaxies. \citet{mulchaey00} noted that previous X-ray telescopes did not have the requisite spatial resolution to separate the intragroup gas from the galaxy-linked emission; however, the resolution of {\em Chandra} is well suited to this task. Futhermore, we join \citet{tamburri12}, who presented a {\em Chandra} study of HCG~79 (also known as ``Seyfert's Sextet''), in speculating that perhaps many compact groups exhibit galaxy-linked X-ray emission rather than a hot intragroup medium associated with the environment. Throughout the remainder of the paper, unless explicitly stated, we discuss the galaxy- and environment-linked diffuse X-ray emission together as they fit into a larger picture of group evolution, i.e., from dynamically unevolved systems with X-ray emission confined to the galaxies to more evolved systems with a single X-ray halo. We stress that it is difficult to disentangle the galaxy- and group-linked emission in the data as this distinction is not always clear, as in the emission surrounding HCG~42A.

\begin{figure}[ht!]
\centering
\includegraphics[width=4.25in]{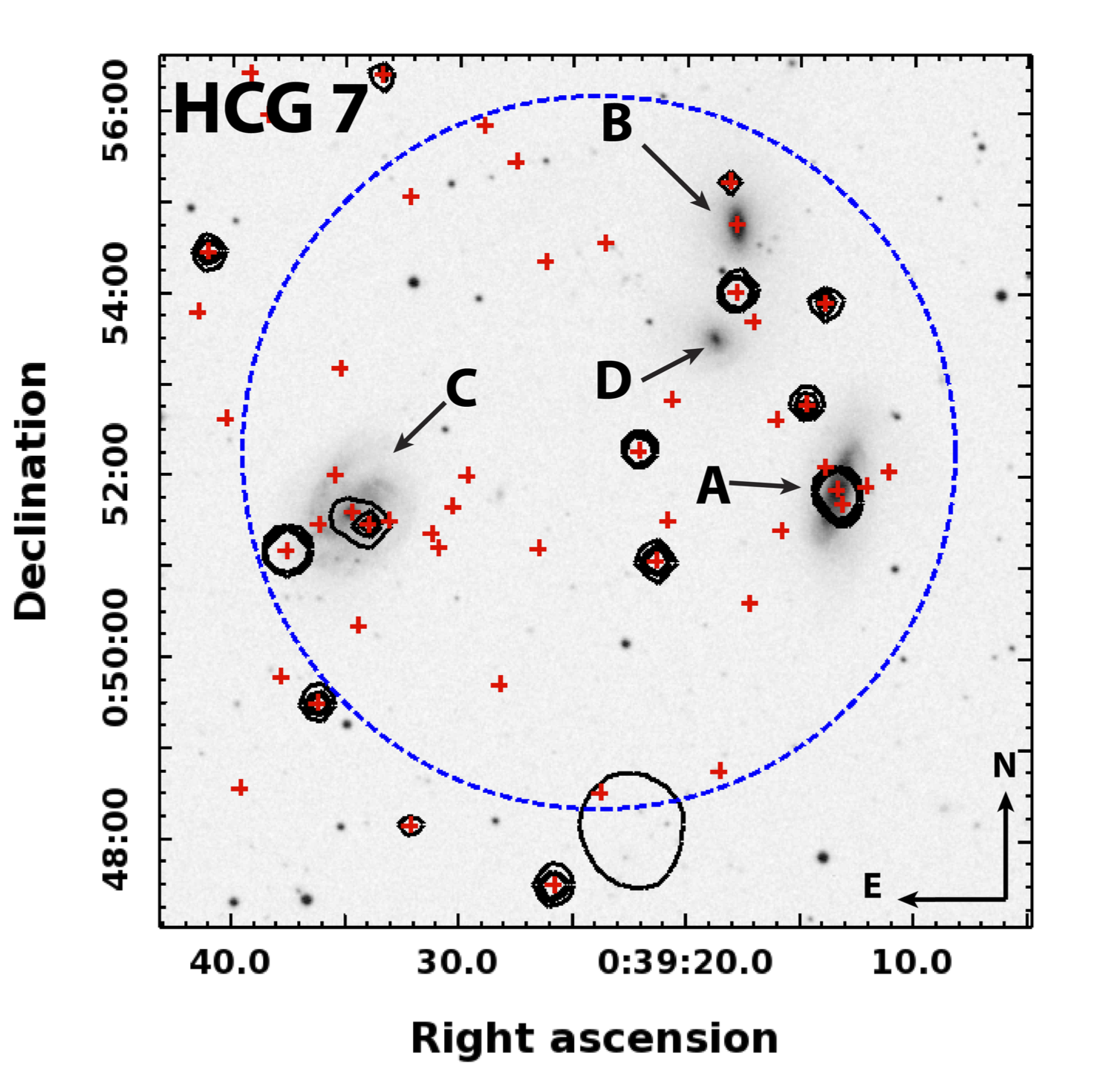}
\caption{Smoothed 0.5--2.0~keV X-ray contours of groups in our sample. The event files were smoothed using the CIAO task {\tt csmooth} with the minimum significance set to 3$\sigma$ above the stowed background scaled by the 9--12~keV count rate. The minimum and maximum smoothing scales used were 2 and 10 pixels, respectively. In all images, the dashed blue line and red crosses correspond to the extraction region used (see Appendix~\ref{app:sourcenotes}) and the locations of the X-ray point sources that were excised from the analysis (but retained in the smoothed X-ray data), respectively. The optical images are DSS POSS2 survey red filter data. Contour levels in counts follow in the captions for each group. {\bf HCG~7:} 0.1, 0.2, 0.3.\label{fig:maps}}
\end{figure}
\clearpage

\addtocounter{figure}{-1}
\begin{figure}[ht!]
\centering
\includegraphics[width=4.25in]{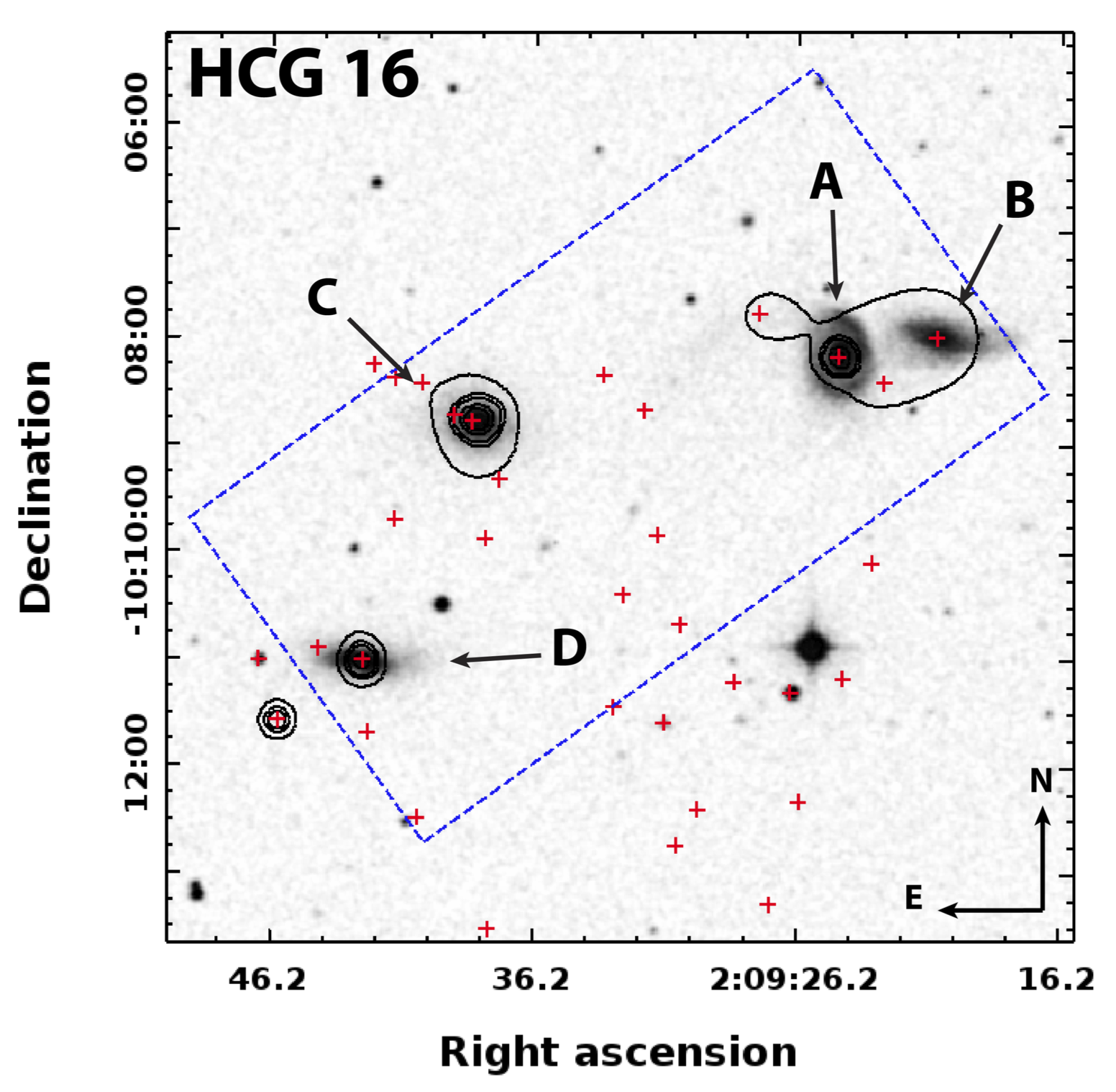}\\
\includegraphics[width=4.25in]{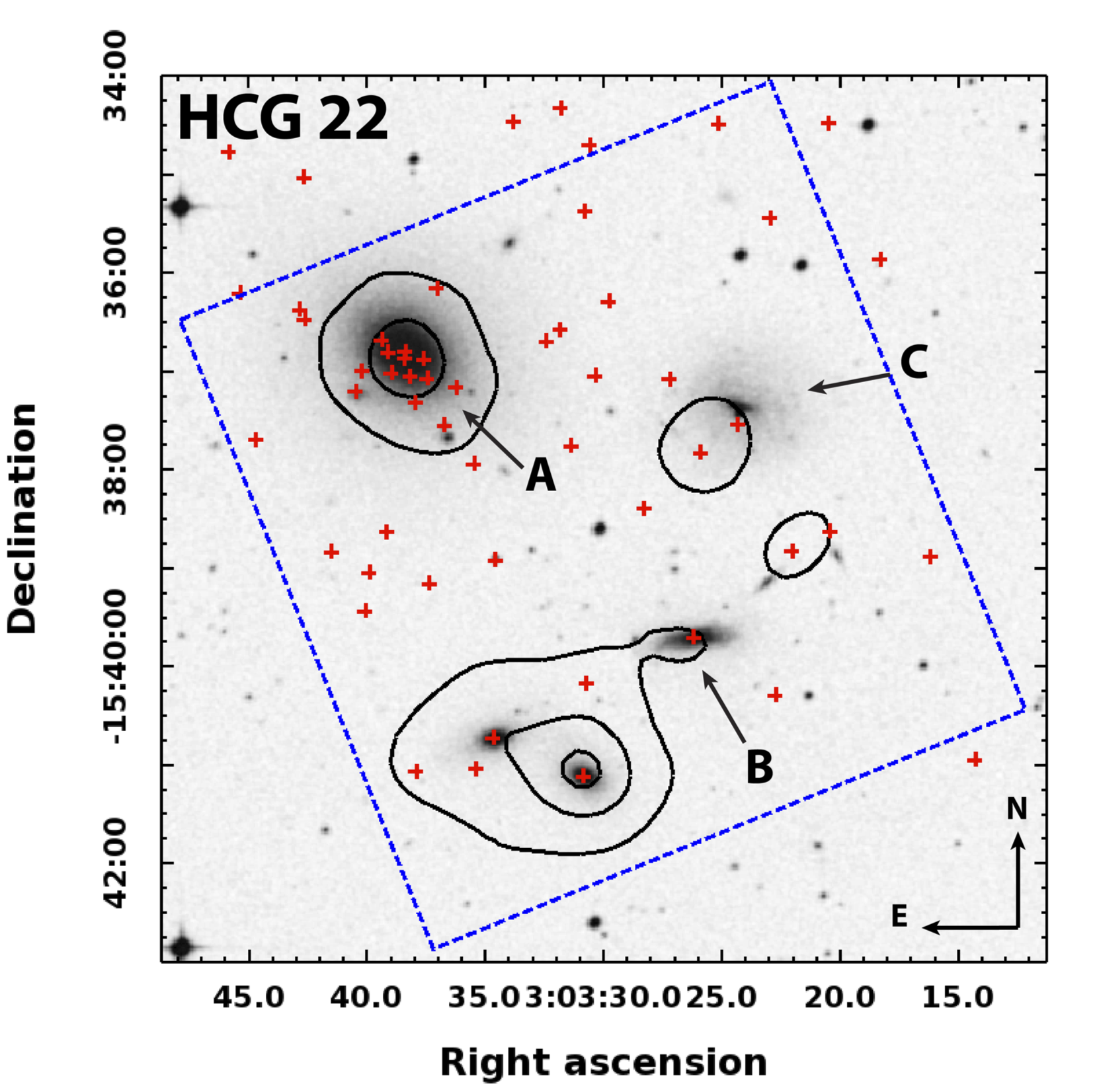} 
\caption[]{Continued. {\bf HCG 16:} 0.1, 0.5, 1, 3, 5, 10. {\bf HCG~22:} 0.1, 0.2, 0.3.}
\end{figure}
\clearpage

\addtocounter{figure}{-1}
\begin{figure}[ht!]
\centering
\includegraphics[width=4.25in]{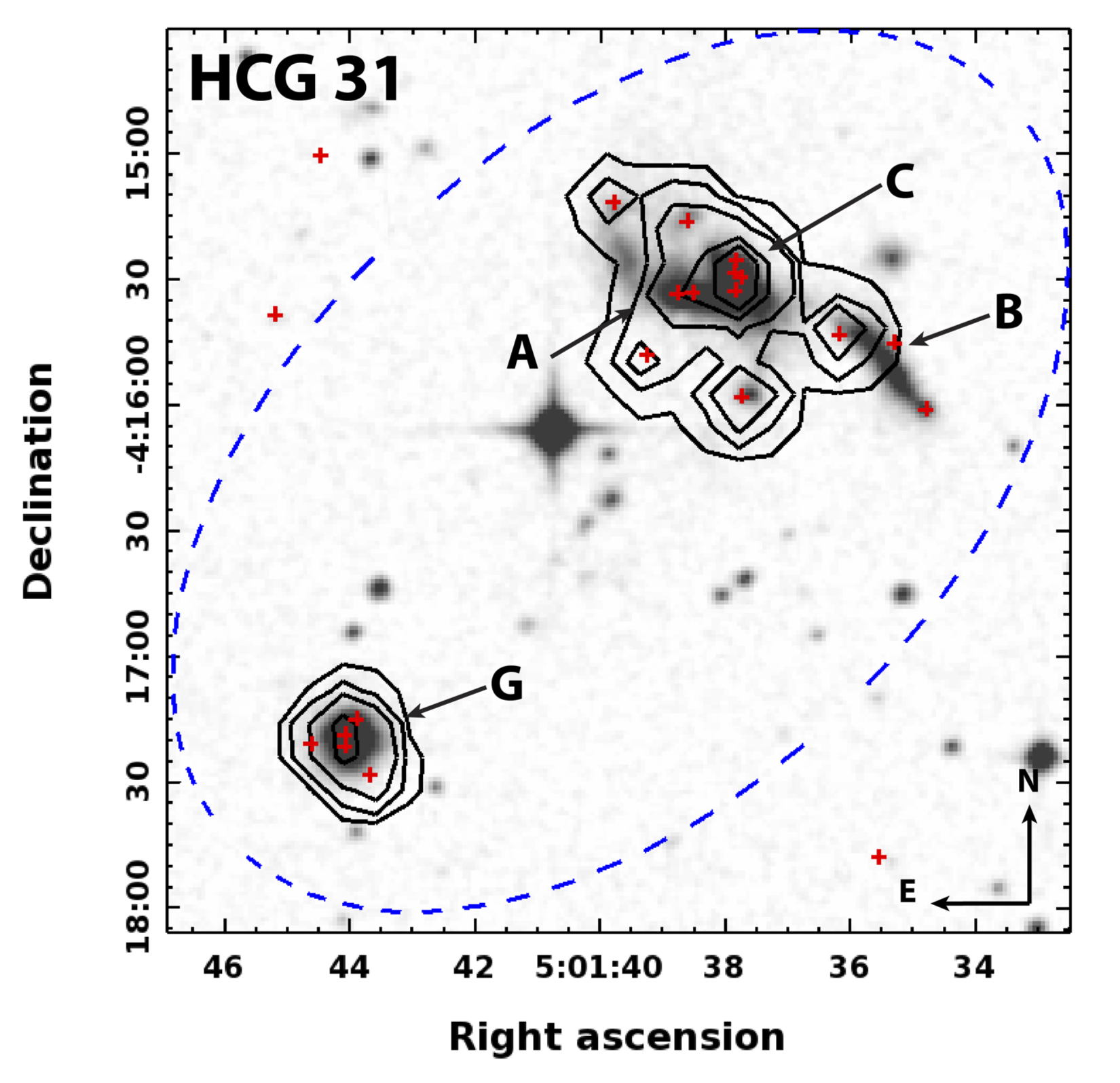}\\ 
\includegraphics[width=4.25in]{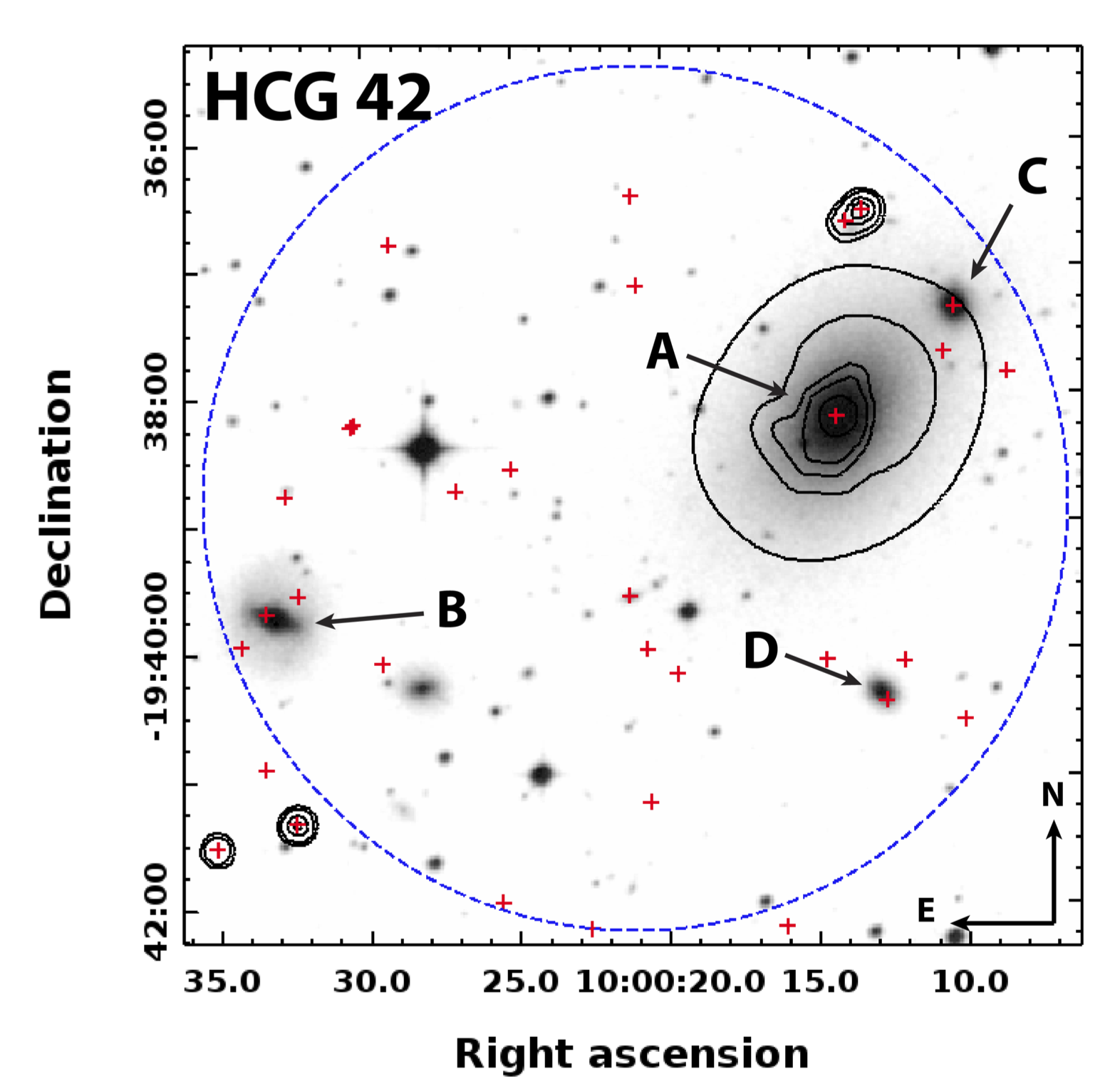}
\caption[]{Continued. {\bf HCG~31:} 0.2, 0.5, 1, 3, 5. {\bf HCG~42:} 0.5, 1, 3, 5, 10.}
\end{figure}
\clearpage

\addtocounter{figure}{-1}
\begin{figure}[ht!]
\centering
\includegraphics[width=4.25in]{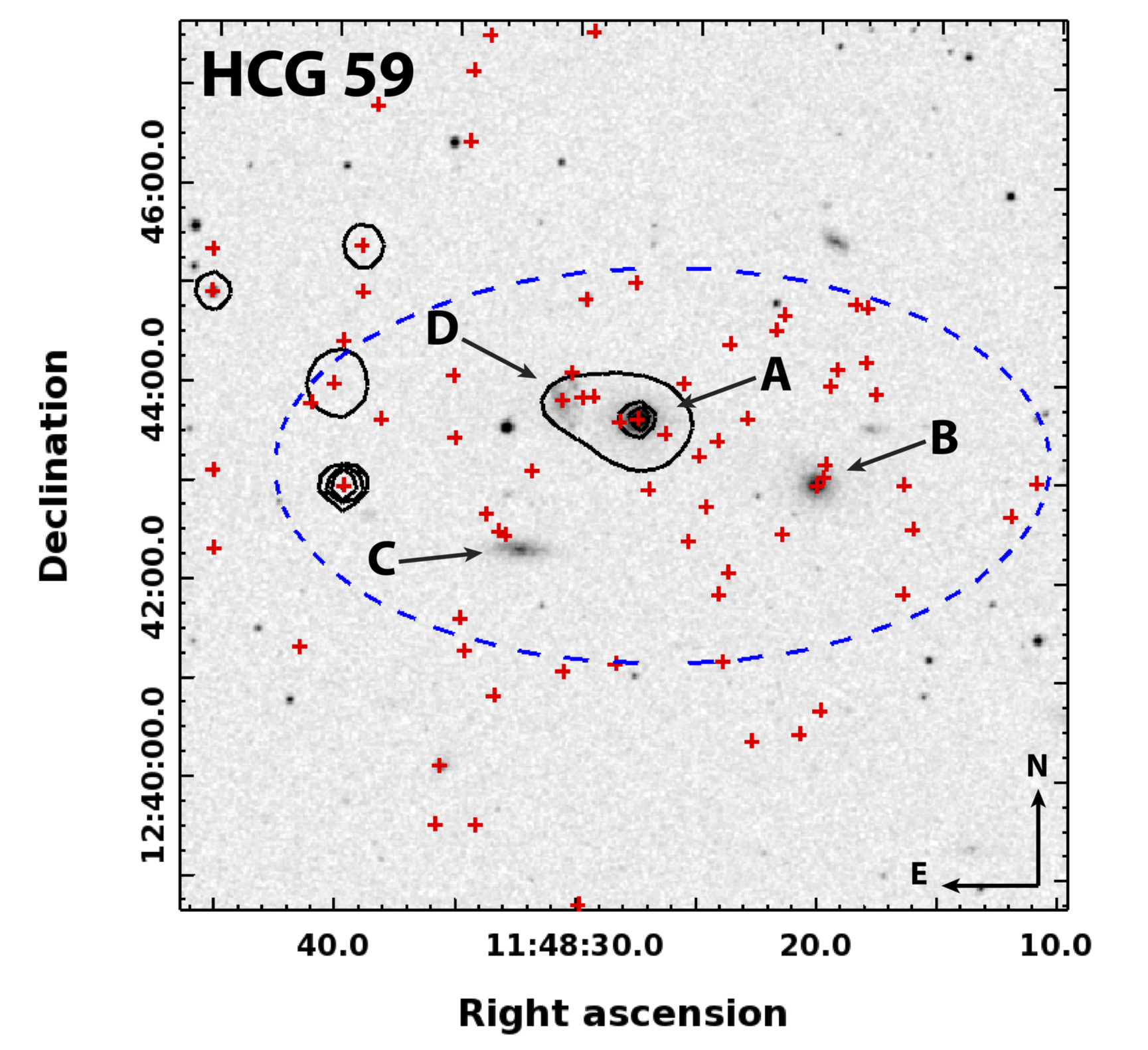}\\
\includegraphics[width=4.25in]{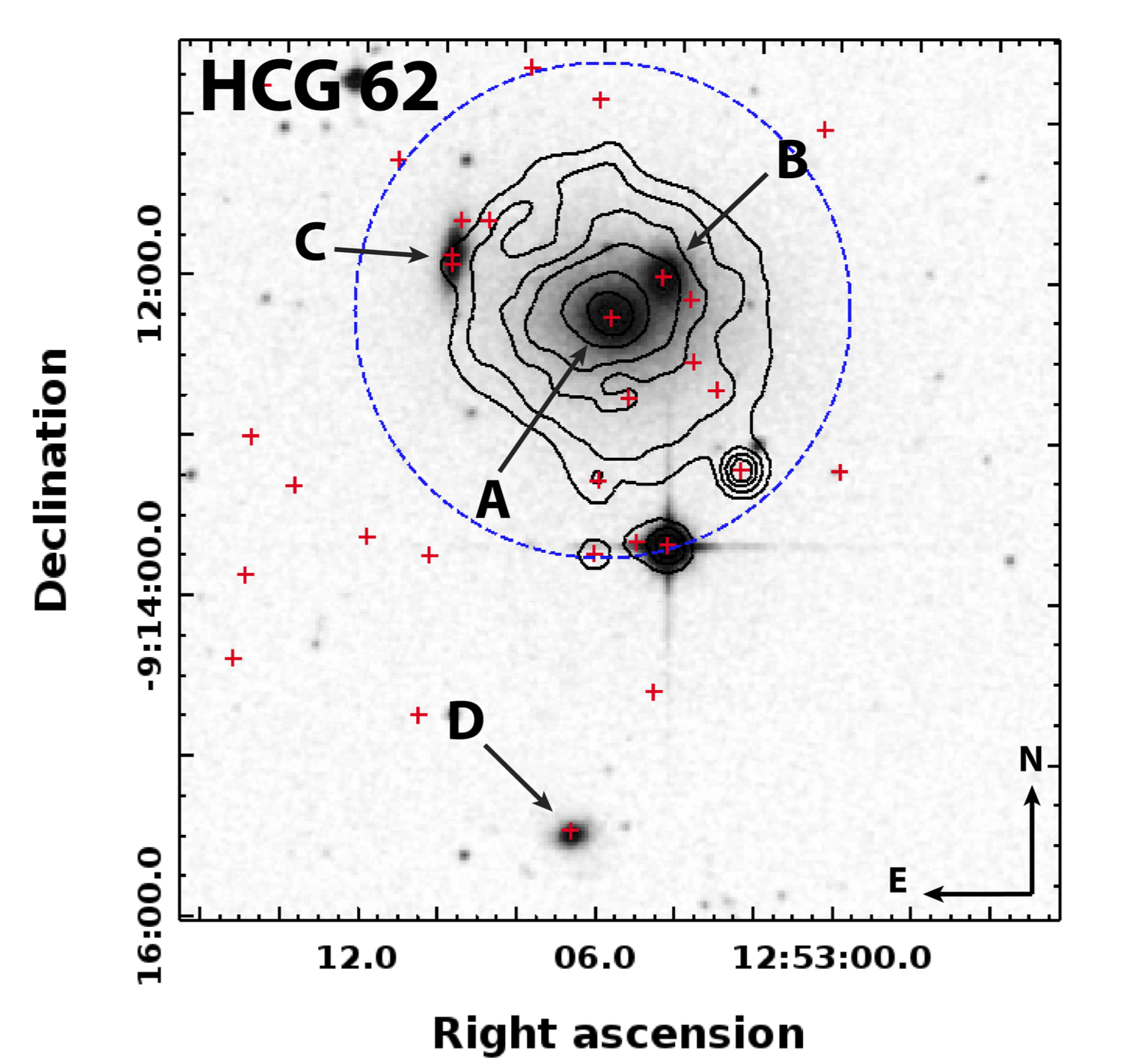}
\caption[]{Continued. {\bf HCG~59:} 0.2, 0.5, 1, 3, 5, 10. {\bf HCG~62:} 3, 6, 10, 15, 30, 50.}
\end{figure}
\clearpage

\addtocounter{figure}{-1}
\begin{figure}[ht!]
\centering
\includegraphics[width=4in]{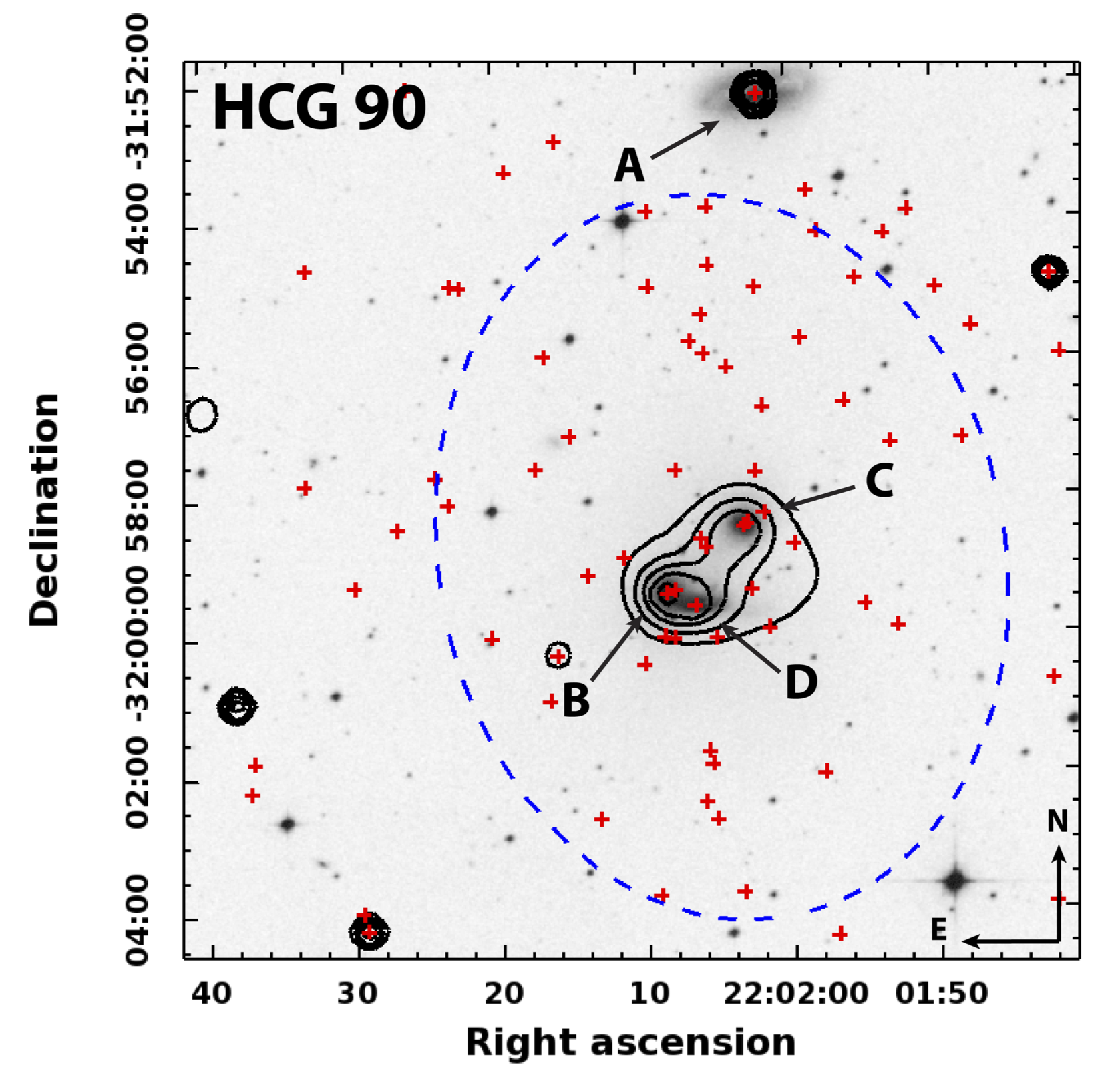}\\
\includegraphics[width=4in]{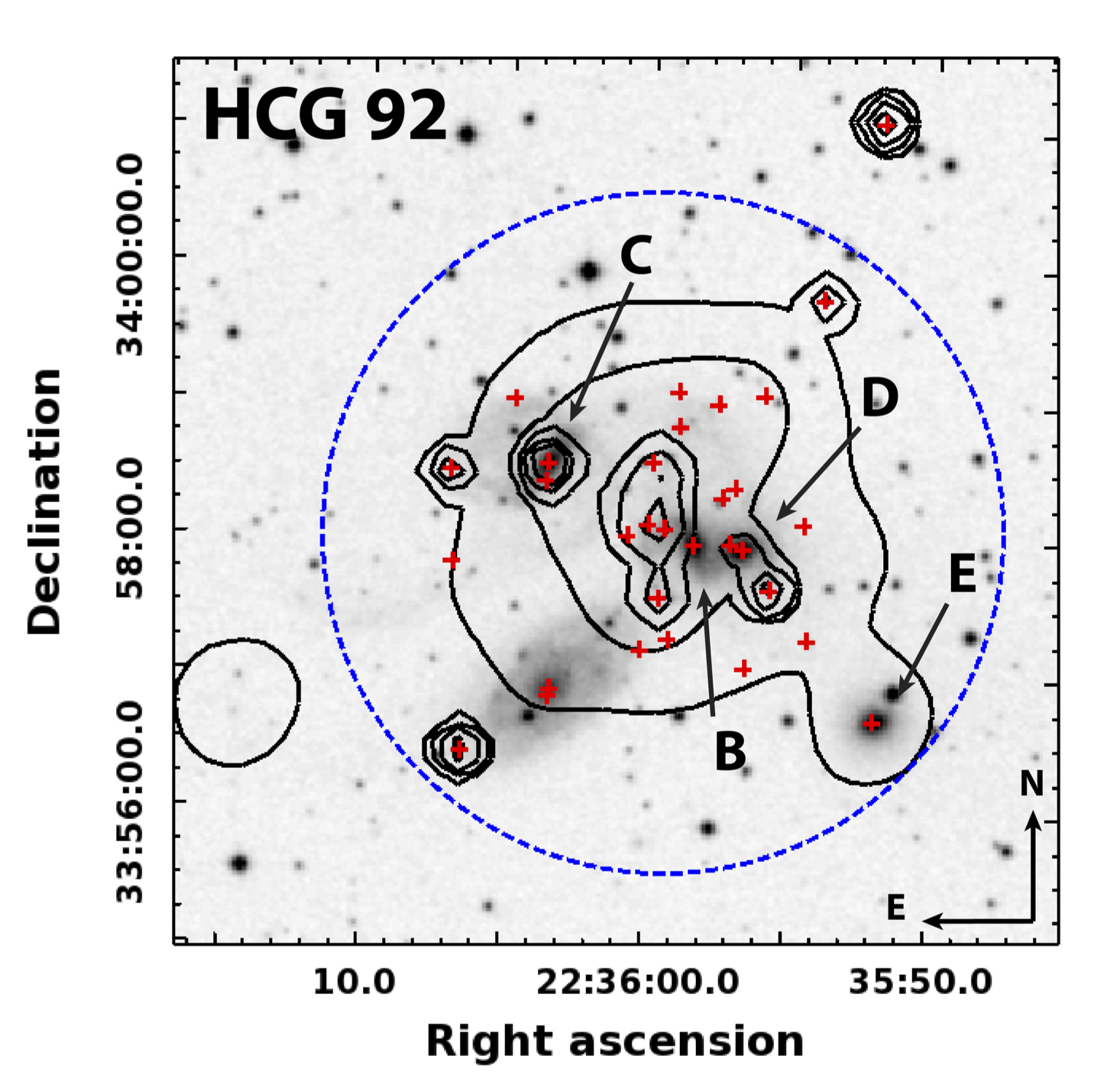}
\caption[]{Continued. {\bf HCG 90:} . {\bf HCG~92:} 0.5, 1, 2, 5, 10.}
\end{figure}
\clearpage
.png

\begin{figure}[ht!]
\centering
\includegraphics[width=\linewidth]{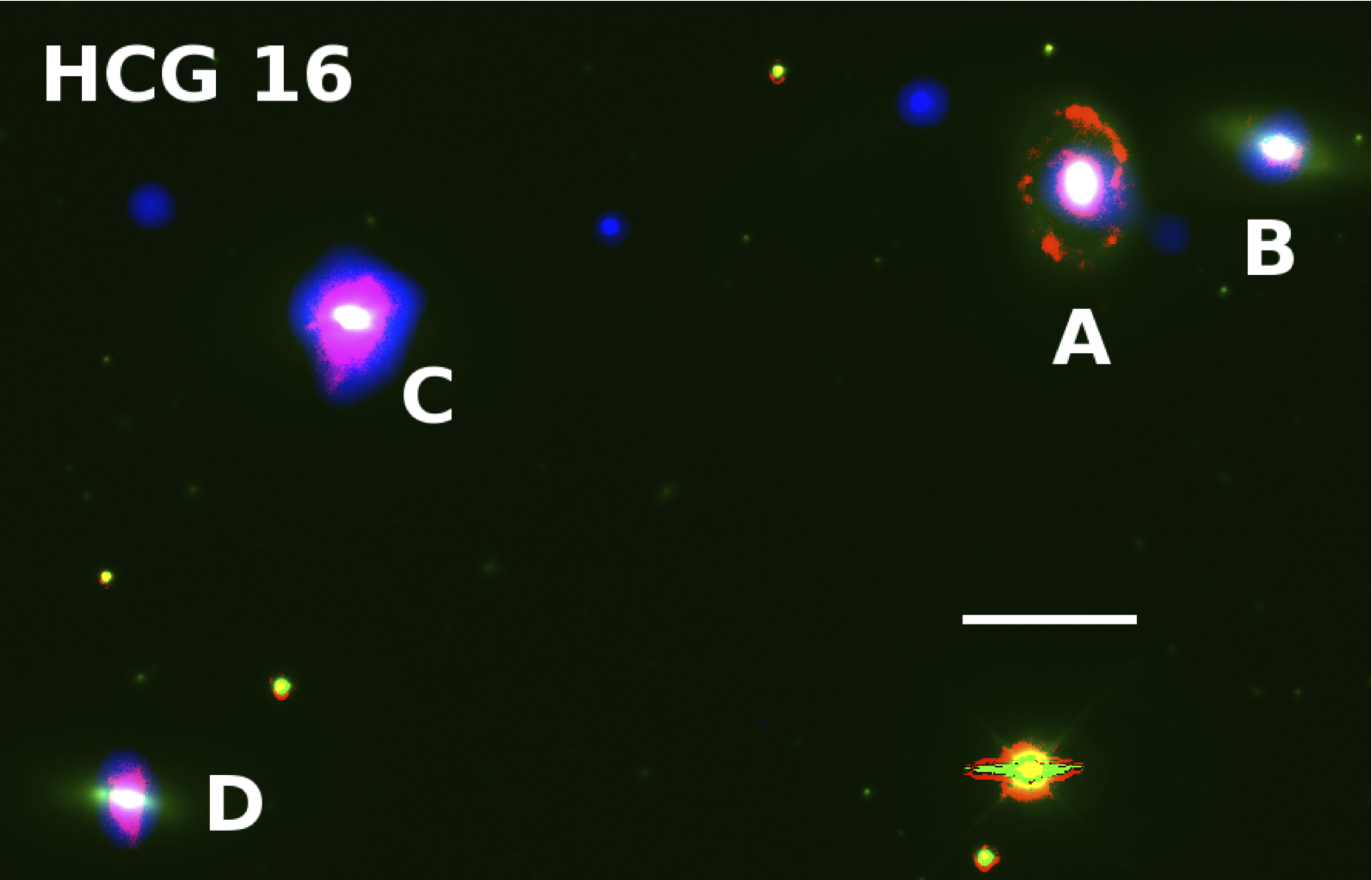}
\caption{Three color image of HCG~16. Red and green represent H$\alpha$ and $R$-band data from CTIO \citep{meurer06}, respectively, while blue is the smoothed soft (0.5--2.0~keV) X-ray emission from {\em Chandra}. North is up and East is to the left in the image. The scale bar indicates 1\arcmin. The pink color in the nucleus of galaxy A and surrounding galaxies C and D is caused by the overlapping of H$\alpha$ and X-ray emission. Galaxy D exhibits a biconical gas distribution centered on its nucleus, while ionized gas surrounds galaxy C with an elongation along the North-to-South axis.\label{fig:hcg16_xalpha}}
\end{figure}

In the specific case of HCG~16, the group contains two members (C and D) with hot gas plumes that are coincident with extended H$\alpha$ emission (see Figure~\ref{fig:hcg16_xalpha}). Both galaxies show clear signs of recent interactions (e.g., disturbed velocity fields; gas/stellar disk misalignments; \citealt{claudia98}; a common \hi envelope surrounding the group; \citealt{verdes01}) and are strong starbursts with SFRs of 14 and 17~\Msun~\per{yr}, respectively, and SSFRs of \e{(2-3)}{-10}~yr$^{-1}$ \citep{ribeiro96,tzanavaris10}. \citet{rich10} used integral field spectroscopy to demonstrate that the motions of the gas and the optical emission-line profiles of 16D are indicative of an M82-like superwind due to the intense starburst within the galaxy, and that 16D and M82 share some similar properties (e.g., metallicity, $L_{\mathrm{IR}}$, $L_{\mathrm{H}\alpha}$). Based on its high star formation rate and extended emission-line gas, 16C may represent a second M82-like starburst in the HCG~16 group, possibly triggered by a recent ($\lesssim1$~Gyr) tidal interaction with 16A similar to the encounter between M81 and M82 $\sim$220~Myr ago \citep{gottesman77,konstantopoulos09}. Furthermore, \citet{jeltema08} found evidence for a weak soft X-ray bridge connecting galaxies 16A and B indicating a recent tidal interaction between these group members as well.

Further examining the morphology of the hot gas and its relation to the galaxies in the compact groups, we find that the gas in low dynamical mass ($\lesssim10^{12.5}$~M$_\odot$) groups with low velocity dispersions ($<250$~km~s$^{-1}$) is concentrated around the individual group members, while gas in relatively higher mass systems with larger velocity dispersions begins to resemble a common envelope.

\subsection{The $L_X-T$ and $L_X-\sigma$ Relations Re-visited}
\label{subsec:scaling}

Previous work has examined the relationships between the X-ray luminosity and both the plasma temperature and the cluster/group velocity dispersion in systems of galaxies (e.g., \citealt{ponman96,mulchaey98,wu99,helsdon00,osmond04,mulchaey03}). The cluster data have been found to show very little scatter in $L_X-T$ and $L_X-\sigma$ space. However groups, with fainter X-ray luminosities, have been found to exhibit a larger spread potentially due in part to uncertainties in the measurements, or because they are not virialized systems (see below). We merge the cluster data from \citet{wu99} and \citet{zhang11} for comparison to the CGs. Prior to merging, we first adjusted the \citet{wu99} data to our assumed cosmology, while \citet{zhang11} used the same cosmology as that assumed in this paper. When comparing the cluster data to the compact groups in $L_X-T$ space, we only include the 176 clusters with uncertainties in both $L_X$ and temperature. This same criterion, with respect to velocity dispersion rather than temperature, is applied to the cluster data when comparing the groups and clusters in $L_X-\sigma$ space, resulting in 142 clusters in this sample. We note that the \citet{zhang11} clusters are measured to $r_{500}$ (i.e., the radius at which the average density falls to 500 times the critical density at that redshift), and the extraction regions used in this work vary between 26\% and 69\% of $r_{500}$. It is unclear to what radius the \cite{wu99} clusters were measured, however the authors state that they used a $\beta$ model to correct all of their data to the same fraction of the virial radius.

\begin{figure}[ht!]
\centering
\includegraphics[scale=0.7]{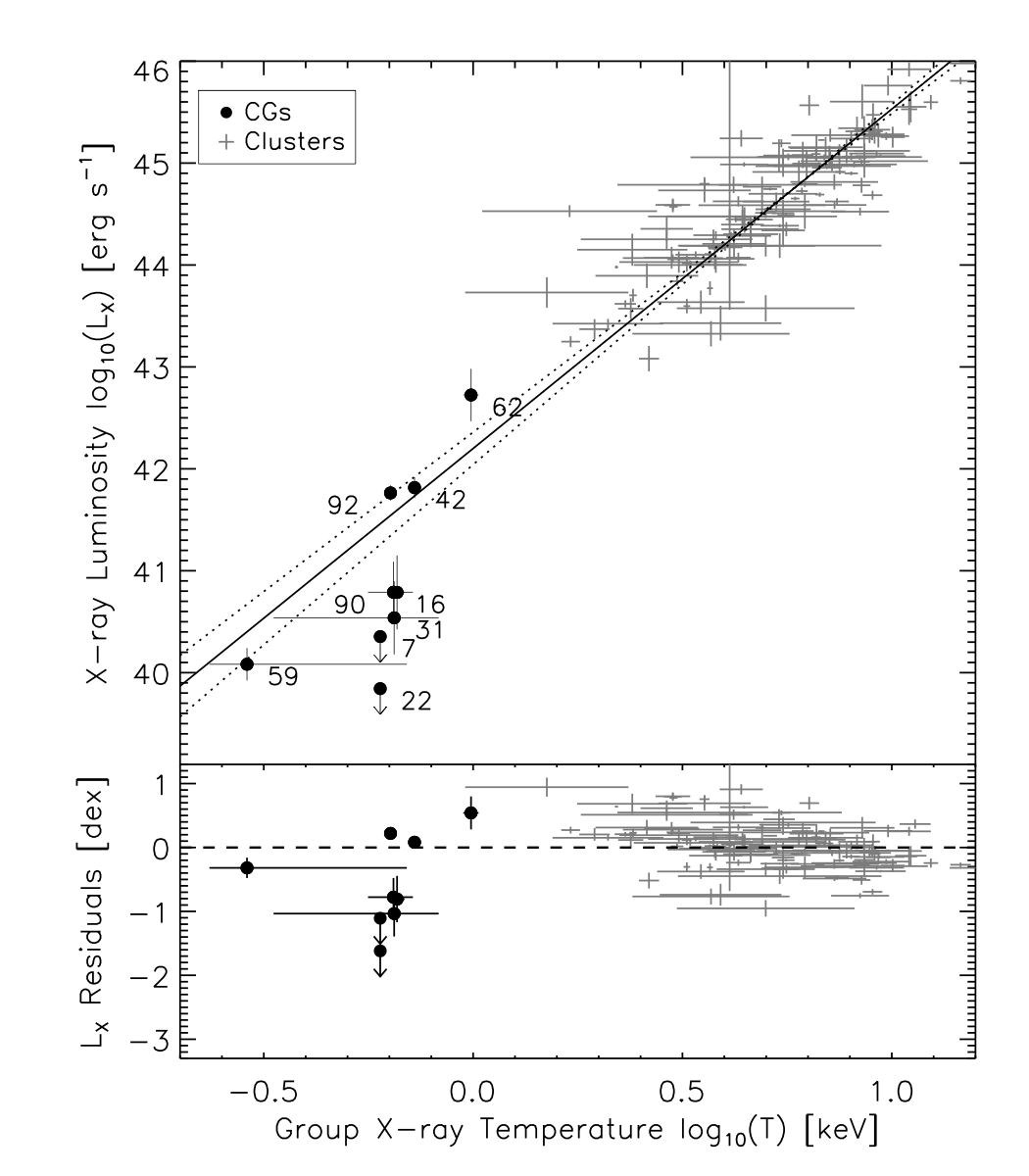}\caption{The $L_X-T$ diagram for compact groups (black filled circles). The solid line indicates the best fit to the cluster data (gray crosses) from \citet{wu99} and \cite{zhang11} using an orthogonal distance regression (ODR) fit to 176 clusters, while the dashed lines represent the errors on the fit. We plot the two plasmas in HCG~62 as one data point using the luminosity-weighted average of the temperatures from Table~\ref{tab:ponmancompare} and the total X-ray luminosity. The temperature of HCG~59 is very uncertain due to the poorly defined peak in the X-ray spectrum, and we conservatively estimate it to be $<$1~keV, however we include the best-fitting temperature from the MEKAL model here in the $L_X-T$ diagram for qualitative purposes. The best fitting cluster relation agrees with HCGs~31, 42, 59, and 92 within the errors, while HCG~62 agrees lies within the scatter in the cluster data. We do note most of the compact groups lie systematically below the cluster fit.\label{fig:kT_v_Lx}}
\end{figure}

Figure~\ref{fig:kT_v_Lx} shows the $L_X-T$ relationship for the groups and the clusters from \citet{wu99} and \citet{zhang11}. Using linear regressions described by \citet{akritas96}, we fit the clusters in $L_X-T$ space {\em not} including the compact groups. For this fit, we use the orthogonal distance regression (ODR) fitting method, which finds $\log_{10}(L_X)~=~(42.2\pm0.2)+(3.33\pm0.20)\log_{10}(T)$, where $L_X$ and $T$ are in units of erg~s$^{-1}$ and keV, respectively. Examining Figure~\ref{fig:kT_v_Lx}, we find that HCGs~31, 42, 59, and 92 agree with the $L_X-T$ cluster relation within the errors. We note that HCG~62 lies slightly above the the cluster relation (i.e.,~it is brighter for its temperature). HCGs~7, 16, 22, and 90 lie below the $L_X-T$ relation from the clusters. 

%The best-fitting line from \citet{wu99} to 142 clusters using an orthogonal distance regression (ODR) method is plotted for comparison. We chose to compare the CGs to the ODR fit, rather than the least-squares regression that \citet{wu99} also performed, because the authors stated that the ODR method gives more robust results. The cluster fit is given as ${L_X}=10^{-0.92\pm0.05}T^{2.72\pm0.05}$ (\citealt{wu99}, equation~3), where $L_X$ and $T$ are in units of $10^{44}$~erg~s$^{-1}$ and keV, respectively. The groups, with one exception, clearly lie well below the cluster relationship, well outside the errors, and therefore do not follow the same scaling law with temperature. We note that HCG~62 does agree with the cluster $L_X-T$ scaling relation if we consider the average temperature and total luminosity. The overprediction of the remaining group X-ray luminosities by the cluster $L_X-T$ relationship indicates that galaxy groups are not fully relaxed systems in which the gas has become virialized as in galaxy clusters, rather non-gravitational processes may be responsible for heating the gas (see below). Examining Figure~\ref{fig:kT_v_Lx}, the groups with brighter diffuse X-ray emission come much closer to the $L_X-T$ fit to the cluster data; therefore it is possible that the groups may agree better with the fit as they dynamically evolve.

\clearpage

\begin{figure}[ht!]
\centering
\includegraphics[scale=0.7]{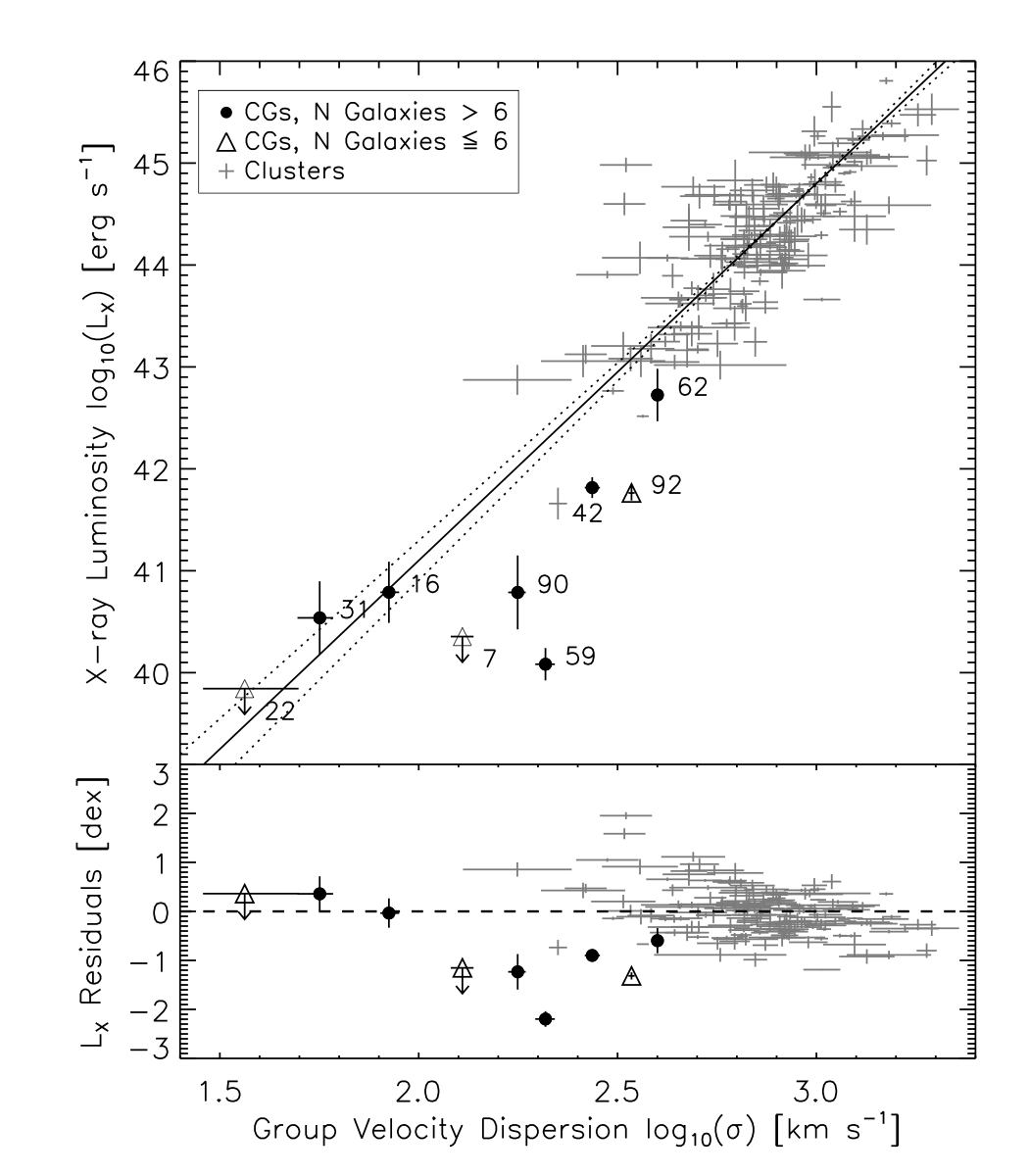}\caption{The $L_X-\sigma$ diagram for compact groups. Due to the inaccuracy in the velocity dispersions for systems with a small number of components, we plot filled black circles to indicate groups for which 6 or more galaxies were used to compute the velocity dispersion, while open triangles represent groups below this threshold. The solid line indicates the best fit to the clusters using a linear-least squares fit to the 142 clusters taken from \citet{wu99} and \citet{zhang11} (gray crosses), while the dashed lines represent the errors on the fit. The best fitting cluster relation agrees with HCGs~16, 31, and 62, while HCGs~7, 42, 59, 90, and 92 lie below the fit. We note that the agreement of HCGs~16 and 31 with the cluster fit is likely coincidental because the emission in these systems is largely due to star formation rather than virialization, therefore they should not necessarily agree with the fit to the cluster data.\label{fig:sig_v_Lx}}
\end{figure}
\clearpage

Figure~\ref{fig:sig_v_Lx} shows the $L_X-\sigma$ relation for the groups in our sample. Velocity dispersions were calculated using the most accurate velocities available from NED for the CG members including additional known group members primarily from the large spectroscopic surveys of \citet{decarvalho97}, \citet{zabludoff98}, and \citet{zabludoff00} (see Table~\ref{tab:targets} for full velocity references). Velocity errors for individual group members were typically $\lesssim30$~km~s$^{-1}$. Again, we compare our compact group data with the cluster data taken from \citet{wu99} and \citet{zhang11} using a linear least-squares fit. We chose this type of fit because the ODR method used in the $L_X-T$ diagram showed strong systematics in the residuals of the cluster data. The resulting best fit is $\log_{10}(L_X)~=~(33.7\pm0.6)+(3.70\pm0.21)\log_{10}(\sigma)$, where $L_X$ and $\sigma$ are in units of erg~s$^{-1}$ and km~s$^{-1}$. We find that two of our CGs (HCGs~16, 31) agree with the $L_X-\sigma$ relation from the galaxy clusters within errors, while HCG~62 is similar to the clusters within the scatter. In the cases of HCGs~16 and 31, this agreement is likely coincidental because the $L_X-\sigma$ relation is predicted from the Virial theorem to be $L_X\propto\sigma^4$, and the bulk of the hot gas in these systems is clearly due to star formation. We do note, however, that for groups in which dynamical processes are increasingly important in heating the gas (i.e.,~HCGs~42, 59, 62, 90, and 92), there does appear to be a monotonic increase in the X-ray luminosity with increasing velocity dispersion, albeit with some scatter.

\citet{ponman96} hypothesized that higher temperatures for a given $L_X$ than would be expected from the cluster $L_X-T$ relation for compact groups could be a result of the injection of energy into the IGM from galactic winds, while \citet{ponman99} suggested that the observed deviation of the galaxy group X-ray luminosities from the established cluster scaling relations could be explained by preheating of the IGM by supernovae.  In the preheating model, heating of the gas in the IGM occurs early in the lifetimes of groups and similarly steepens the $L_X-T$ relation for these systems. Despite groups such as HCG~16 where star formation may play a more prominant role, the existence of X-ray brighter groups (i.e., HCGs~42) below the $L_X-T$ scaling relation for clusters indicates that galaxy winds alone are unlikely to explain the observed deviation. However, the observed dichotomy in X-ray gas morphology (i.e., galaxy- vs. environment-linked emission) may indicate that the low-mass, low velocity dispersion systems are dynamically unevolved and have shallow potential wells that are unable to heat any neutral gas that has been liberated from the group members. In these systems, the role of individual galaxies may be more important in heating gas through local (e.g., star formation, superwinds, and accretion) rather than global processes (e.g., virialization). As systems accrete more mass or as the system relaxes, the potential well will deepen and the contribution of individual group members to the diffuse X-ray emission should lessen. 

After attempting to subtract the X-ray emission associated with the group members in HCGs, \citet{ponman96} found that the $L_X-\sigma$ relation was flattened compared to the clusters (i.e., the rate of change in $L_X$ as a function of $\sigma$ was slower for groups than for clusters), similar to the \citet{dellantonio94} {\em Einstein} study of rich groups\footnote{While \citet{dellantonio94} did not subtract the galaxy-linked emission from their data, they did avoid using data in instances where the hot gas was clearly associated with only galaxies and not the group environment.}, as well as a subsequent study of CGs by \citet{helsdon00}. The morphology of the hot gas in CGs may provide an explanation for the groups that disagree with the $L_X-T$ and $L_X-\sigma$ relations derived from galaxy clusters. As discussed in \S\ref{sec:morph}, the hot IGM that permeates HCG~62 and the X-ray bright halo centered on HCG~42A have morphologies that are qualitatively similar to the hot gas observed in the cluster environment, i.e., where the gas has been heated by virialization. In the remaining groups, the only diffuse X-ray emission we detect is associated with the individual galaxies rather than the group environment itself. This suggests that the gas is not in hydrostatic equilibrium within the group, and therefore the temperature and luminosity of the total X-ray gas in the system does not trace the group potential as it does in the galaxy clusters. %In the case of $L_X-\sigma$ for individual galaxies, the known relationship for ellipticals has a steeper slope compared to that of clusters (e.g., $L_X\propto\sigma^m$, where $m=4.4$ or 10.2 for clusters and ellipticals, respectively; \citealt{mahdavi01}). In fact, given that the velocity dispersions of compact groups are similar to those of individual galaxies, it is possible that in some cases we are probing the $L_X-\sigma$ relation for galaxies (e.g.,~HCG~42, in which the X-ray emission may be the halo of just 42A) rather than the relation for the group environment. The steeper slope observed in the $L_X-\sigma$ relation for ellipticals could explain the X-ray brighter groups HCGs 42 and 62; however, the fainter groups (e.g.,~HCG~31) appear to follow a different relation. A strict analysis of the $L_X-\sigma$ scaling relation in groups requires further data to study robustly the individual group members against a sample of non-group galaxies with similar characteristics (e.g., morphology) and fully examine the impact of the compact group environment on the hot gas properties.

For completeness, we compare our sample of compact groups to studies of the X-ray scaling relations in both normal (i.e.,~those that are not compact) and fossil groups. With respect to the $L_X-T$ relationship, \citet{khos07} and \citet{harrison12} noted that fossil groups match well with both clusters and normal groups. Only three of the compact groups presented in our study (HCGs~42, 62, and 92) agree well with the cluster scaling relation within the scatter, though HCG~92 is dominated by emission from a strong shock. The remaining CGs at $T\lesssim0.65$~keV fall below the best fit to the cluster data. In $L_X-\sigma$ space, \citet{rines10} showed that normal groups did not deviate from the relation for clusters, while \citet{khos07} found that fossil groups are consistent with the best fit to the cluster data, though more X-ray luminous than the normal groups in their sample. Our sample of compact groups fall systematically below the observed $L_X-\sigma$ relation for clusters with the exceptions of HCGs~16 and 31, which are dominated by vigorous star formation. This may be in agreement with the interpretation of merging systems being X-ray underluminous for their velocity dispersions (e.g.,~\citealt{rasmussen06,popesso07}). We caution the reader that in most cases the velocity dispersions of the groups are measured from very few galaxies, therefore they likely do not accurately represent the three dimensional dispersions that are assumed by the $L_X-\sigma$ relation. %We do note that five of the X-ray detected groups fall systematically below the best fit to the cluster data, while the two groups that are consistent with the $L_X-\sigma$ relation have emission that is dominated by vigorous star formation.

%For completeness, we compare our sample of compact groups to studies of the X-ray scaling relations in both normal (i.e.,~those that are not compact) and fossil groups. In terms of the $L_X-T$ relation, \citet{khos07} and \citet{harrison12} noted that fossil groups match well with hot clusters and the cooler, normal groups; however, the former study does indicate that the normal groups do deviate from the cluster relation and contain more scatter at low temperatures. \citet{eckmiller11} found that the normal groups did not significantly deviate from the $L_X-T$ cluster relation until they imposed a cut in the temperature at 3~keV. Below this cut, the normal groups follow a statistically steeper relation than do the clusters. In $L_X-\sigma$ space, \citet{khos07} showed that the fossil groups also line up well with both the normal groups and clusters, while \citet{rines10} demonstrated that the scaling relation for X-ray selected groups did not deviate from those of higher mass clusters or lower mass normal groups. \citet{rines10} did caution that recent studies (e.g., \citealt{rasmussen06,popesso07}) indicate that dynamically young systems are X-ray underluminous for their velocity dispersions.  

Based on the discrepancies between the compact groups and the X-ray cluster scaling relations, we postulate that systems similar to the {\em low-mass, X-ray faintest} groups in our study should not be considered analogs to clusters, with the possible exception of HCG~62, which lies close to the cluster data in both $L_X-T$ and $L_X-\sigma$ within the scatter. It is important to note that this dissimilarity between the low-mass groups and clusters does not preclude these systems from becoming more cluster-like if they somehow become similar to the more massive, rich groups in our sample (e.g., by accreting additional members and continuing strong interactions to liberate gas into the IGM). However, it is unlikely that rich groups today formed from poor groups like those observed in the current epoch. Therefore the exact mechanism whereby the poor groups in our sample could become more cluster-like remains unknown. We do note, however, that there could be extended, faint emission that is undetectable in the available data (e.g., the X-ray IGM of HCG~16; \citealt{belsole03}). Deeper observations may reveal cooler, lower luminosity gas associated with entire groups rather than the galaxies, which could shed light on the relationship between poor groups and more cluster-like systems.

\subsection{Comparison of the X-ray Data with \hi Gas}
\label{subsec:hi}

Previous work has shown that relatively \hi rich compact groups contain galaxies that exhibit mid-IR colors dominated by star formation with correspondingly high SSFRs measured from UV+24~$\mu$m fluxes, while groups deficient in \hi have more quiescent colors and low SSFRs \citep{johnson07,walker10,walker12,tzanavaris10}. This suggests an evolutionary sequence of compact groups in which the \hi gas is processed either through star formation or ionization by the group potential (the velocity dispersions imply virial temperatures of $\sim$0.08--0.3~keV; \citealt{verdes01,johnson07}). Therefore, we expect that the most \hi abundant groups should have very weak diffuse X-ray emission, while the \hi poor groups should have the brightest X-ray luminosities. Furthermore, the morphology of the \hi gas may dictate how it is processed, i.e., through star formation or by virialization in the IGM in CGs with neutral gas confined to the galaxies or stripped into the IGM, respectively (cf.~\citealt{konstantopoulos10}). Throughout this section, we use the \hi abundance type notation from \citet{johnson07}. Types~I, II, and III indicate decreasing \hi abundance relative to the group dynamical mass, respectively. \citet{johnson07} quantitatively define these \hi mass types as (I) log(M$_{\mathrm{H}~{\textsc{i}}}$)/log(M$_{\mathrm{dyn}})\geq0.9$, (II) $0.9>\log(\mathrm{M}_{\mathrm{H}~{\textsc{i}}}$)/log(M$_{\mathrm{dyn}})\geq0.8$, and (III) log(M$_{\mathrm{H}~{\textsc{i}}}$)/log(M$_{\mathrm{dyn}})<0.8$. The \hi evolutionary types for each group are listed in the last column of Table~\ref{tab:targets}.

\begin{figure}[t!]
\centering
\includegraphics[width=\linewidth]{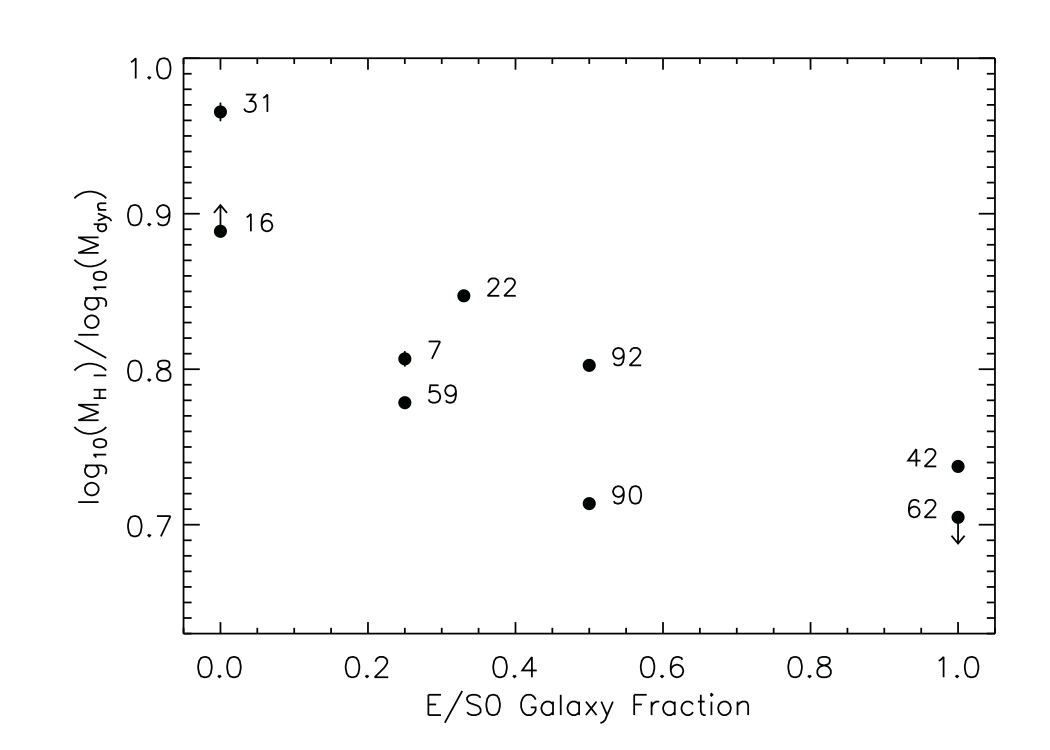}\caption{The \hi gas mass ratio as a function of the main group member morphologies. The \hi mass relative to the group dynamical mass decreases as the fraction of E/S0 galaxies increases. The general trend of the data points indicates that CGs exhaust their \hi reservoirs (i.e., $M_{\mathrm{H I}}\lesssim0.1$\% $M_{\mathrm{dyn}}$) when approximately 50\% of the main group members have E/S0 morphologies. \label{fig:hi_earlytype}}
\end{figure}

We calculate the \hi to dynamical-mass ratios for the CGs in our sample using the most precise velocities available from NED, the two galaxy median separators from \citet{hickson92}, and total group \hi masses from Green Bank Telescope, Arecibo Observatory, and Effelsberg 100~m Antenna single dish measurements by both \citet{verdes01} and \citet{borthakur10}\footnote{The \hi masses of HCGs~16 and 62 are lower and upper limits, respectively. The lower limit of the mass in HCG~16 is due to the large angular size of the \hi envelope, which extends beyond the Green Bank Telescope beam.}. Subtypes based on the morphology of the \hi gas, which we qualitatively assessed from VLA observations, are included as part of the proposed evolutionary sequence of compact groups \citep{konstantopoulos10}. Type~A groups are those in which the neutral gas is confined to the individual group members, while Type~B groups have \hi gas distributed between the galaxies and not centered on any particular member(s). We first test the use of the \hi to dynamical-mass ratio as a descriptor for group evolutionary state by comparing it to the group E/S0 galaxy fraction for the main galaxies. Using the \hi mass to dynamical-mass ratio to characterize the evolutionary state is preferred over the E/S0 fraction because the mass ratio is a continuous distribution, while the E/S0 fraction values are discrete due to the small number of relatively massive members in each group. To quantify the relationship, we used the {\tt ASURV} statistical package \citep{lavalley92}, which implements the methods presented in \citet{isobe86}, to compute the Spearman rank correlation coefficient. This test measures how well the data are fit by a single monotonic function. From Figure~\ref{fig:hi_earlytype}, we find that the \hi mass normalized by the group dynamical mass decreases with increasing E/S0 galaxy fraction (67\% probability from Spearman test). This result is expected if the \hi to dynamical-mass ratio is indeed a tracer of the evolutionary state of the system.

\begin{figure}[t!]
\centering
\includegraphics[width=\linewidth]{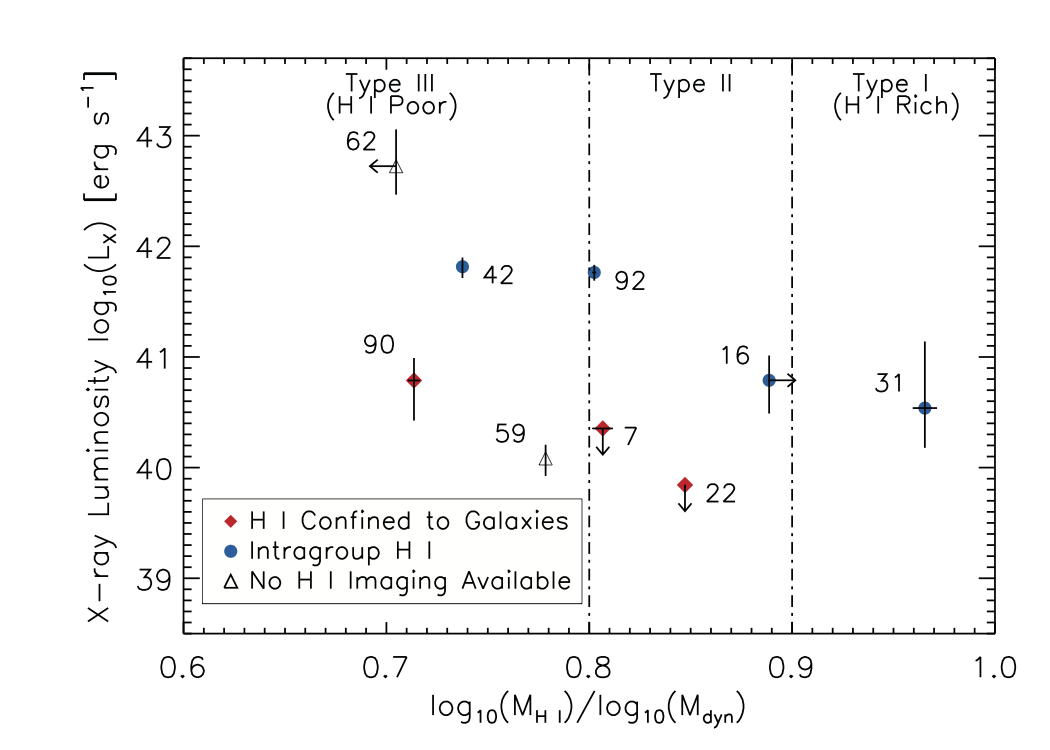}\caption{Distribution of diffuse X-ray luminosity in CGs as a function of the ratio of \hi mass to dynamical-mass used in the evolutionary typing scheme defined by \citet{johnson07}. Neutral gas masses are taken from Green Bank Telescope, Arecibo Observatory, and 100~m Effelsberg Antenna single dish measurements by \citet{verdes01} and \citet{borthakur10}, while the \hi morphologies are qualitatively assessed from VLA interferometric observations, which we lack for HCGs~59 and 62. Note that the values of the \hi mass in HCGs~62 and 16 are upper and lower limits, respectively, and the plotted \hi mass ratio corresponds to these limits. Futhermore, uncertainties in the \hi masses from \citet{verdes01} are not available, therefore we only plot error bars in the \hi mass ratio for data from \citet{borthakur10}. After separating the groups by \hi distribution subtype based on \citet{konstantopoulos10}, there appears to be a distinction in the X-ray luminosity between the two populations. Due to the small number of groups in our sample, more data are required to concretely determine if this difference is real.\label{fig:evolution}}
\end{figure}

When we examine how $L_X$ changes with \hi mass ratio, we observe that the \hi poor Type~III HCGs~42, 59, 62, and 90 are X-ray brighter compared to more \hi rich groups as expected if the gas has been processed by star formation or heated by the group potential. Of the Type~II groups, we note that HCG~92 appears to be nearly as bright as the \hi poor CGs, probably due to the shock front created by the high velocity intruder galaxy in the group (e.g.,~\citealt{trinchieri03,osullivan09}). The only \hi rich Type~I in our study is HCG~31, which contains diffuse X-ray emission linked to star formation activity. Figure~\ref{fig:evolution} shows the X-ray luminosity as a function of the \hi mass ratio. From the figure, we can see that compared to the hot gas in other Type~III groups, HCGs~59 and 90 are 1--2 orders of magnitude fainter than the remaining \hi-poor HCGs. In the case of HCG~90, this could stem from the exclusion of galaxy~A from the X-ray analysis due to its bright Seyfert~2 nucleus (see Appendix~\ref{app:sourcenotes}), however it is unlikely that including the diffuse emission from 90A would increase the total group luminosity by an order of magnitude to bring it to the level of the brighter Type~III groups. 

\begin{figure}[t!]
\centering
\includegraphics[width=\linewidth]{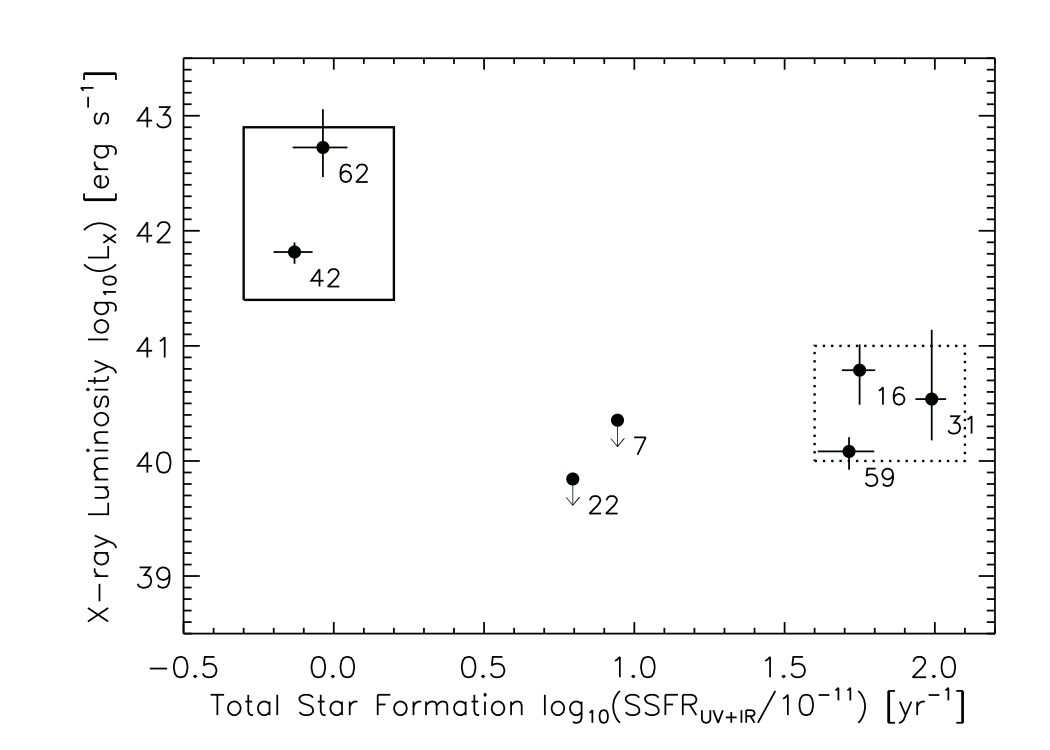}\caption{X-ray luminosity as a function of the specific star formation rate for the seven groups with SSFRs from \citet{tzanavaris10}. The groups primarily fall into two classes: quiescent, X-ray brighter systems (solid box); and star-forming, X-ray fainter systems (dotted box). The boxes are only used to identify the groups in these two regimes, therefore the absolute positions and sizes of the boxes in the figure do not necessarily carry physical meaning. At the low and high SSFRs, the processes that give rise to the X-ray emission are dominated by the group potential and local means (e.g., superwinds), respectively.\label{fig:ssfr_lx}}
\end{figure}

From Figure~\ref{fig:evolution}, we note that if we include the \hi distribution subtypes of the groups, then there may be two distinct trends between the total diffuse X-ray luminosity and the \hi evolutionary type. Specifically, the Type~A groups (\hi confined to galaxies) appear to be consistently less luminous in X-rays compared to Type~B groups (\hi in the IGM). Based on this tentative result, it is reasonable to predict that HCG~62 should have any remaining \hi dispersed throughout its IGM, while \hi gas in HCG~59 should be mostly contained in and around its group members. However, we note that this result is preliminary and that more data are required to quantitatively assess the likelihood that the two \hi gas morphologies are distinct populations in X-ray luminosity. If we assume these \hi morphology classifications for HCGs~59 and 62 to increase the number of data points in each sample to the minimum required, a two sample K-S test gives a 95\% probability that the \hi subtypes are two distinct populations in X-ray luminosity. To examine how our assumption concerning the \hi subtypes of HCGs~59 and 62 influenced the test, we perform a second K-S test in which we switch the \hi morphology classifications for these groups (i.e.,~HCG~59 has intragroup H~{\sc i}, while 62 has \hi only in the galaxies). This results in an 12\% probability that the two \hi morphologies represent distinguishable populations in $L_X$. Due to the small sample size, we cannot determine if this distinction is statistically significant, and further data are required to clarify this phenomenon.

We note that the mix of galaxy morphologies in the groups complicates the comparison of the \hi and X-ray gas. However, \citet{verdes01} and \citet{borthakur10} reported on the observed \hi masses in these groups and the predicted \hi masses based on the group member luminosities and morphologies using relations from \citet{haynes84} who examined \hi in a sample of isolated galaxies. From this, we can compute \hi deficiencies for the groups in our sample, where deficiency describes the ratio of the \hi mass observed to that predicted (in contrast to our use of \hi rich and poor, which describes the relative neutral gas mass normalized to the dynamical mass). Using the nomenclature of \citet{borthakur10} (i.e., ``heavily'' deficient, ``slightly'' deficient, and normal groups contain $<1/3$, between $1/3$ and $2/3$, and $>2/3$ of their predicted H~{\sc i}, respectively) we find that the X-ray fainter groups are either heavily deficient (7, 22, and 90) or normal (59) groups, while the brightest groups are either slightly deficient (16, 31, 62\footnote{This group is likely heavily deficient as its \hi mass is actually an upper limit and the ratio of the limit to the predicted value is 35\% \citep{verdes01}.}, and (92) or normal (42) groups. 

Because the gas is not in hydrostatic equilibrium (as indicated by the diffuse X-ray morphology; see \S\ref{sec:morph}), and in many cases linked to the individual group members rather than the environment, we cannot use the temperature to calculate the group hot gas mass (such a procedure assumes hydrostatic equilibrium in three dimensions). However, if we did assume that the gas is in hydrostatic equilibrium, then the narrow range of temperatures implies a similarly narrow range of masses \citep{fabricant80}. The X-ray contours in Figure~\ref{fig:maps} make this unconvincing due to the extremely large number densities and/or metal abundances necessary to make the hot gas masses of groups such as HCG~16 similar to ones such as HCGs~90 or 62. If we instead hypothesize that the X-ray gas mass is proportional to the bolometric X-ray luminosity, i.e., that there exists a $L_X-M_X$ relation for groups (as in \citealt{zhang11} for clusters), then we can compare the relative X-ray luminosities of the groups as a proxy for their {\em relative} hot gas masses. That the X-ray fainter groups for a given \hi mass ratio are deficient in \hi and that all of the CGs have similar X-ray temperatures suggests two scenarios: either (1) additional X-ray gas has too low surface brightness to detect in the available data, or (2) some fraction of it is missing from the groups. If the \hi was not converted to hot gas, then where is the missing gas in compact groups? The least massive groups in our sample have velocity dispersions that correspond to virial temperatures of only $\sim$0.08~keV, however if the gas is cooler than the virial theorem implies, then it may be in the form of a diffuse UV intragroup medium. Further data are required to fully explore this phenomenon.

\subsection{Diffuse X-ray Emission and Specific Star Formation Rates}

We also compare the X-ray emission of seven groups in our sample to the total group SSFRs for the main group members calculated from the UV+24~$\mu$m fluxes measured by \citet{tzanavaris10} with corrections from Tzanavaris~(2012, private communication) in Figure~\ref{fig:ssfr_lx}, and find that there is a distinction in $L_X$ for detected groups on either side of the SSFR gap (data are not available for HCGs~90 and 92). We exclude dwarf group members with measured SSFRs because HCG~31 is the only CG in our sample that has such measurements of its dwarf population, which increases the total group SSFR by several orders of magnitude due to the combination of relatively low to moderate SFRs and small stellar masses in star-forming dwarf galaxies. In our sample, CGs containing star-forming galaxies all share similar X-ray luminosities and are 1--2 orders of magnitude fainter than groups with low total SSFRs. This may indicate an ``X-ray gap'' analogous to the SSFR and mid-IR color gaps found by \citealt{tzanavaris10} and \citealt{walker10,walker12}. We note that HCGs~7 and 22 have SSFRs that lie within the gap range of $(0.3-1.8)\times10^{-11}$~yr$^{-1}$, and neither has any detected diffuse X-ray emission.

The brighter $L_X$ values associated with the low SSFR groups are due to the advanced evolutionary stage of these environments. HCGs~42 and 62 have very low \hi to dynamical-mass ratios, while simultaneously their relatively massive galaxies are entirely E/S0 types. From the evolutionary scenario presented by \citet{verdes01}, and the fact that compact environments are favorable to the tidal stripping of gas from galaxies \citep{freeland09}, we should expect that these older compact groups have removed the cool gas from the galaxies and heated it in the intragroup medium. Conversely, the star-forming groups HCGs~16, 31, and 59 have relatively faint X-ray emission associated with them. These groups also all exhibit only galaxy-linked emission, as expected from their relatively young evolutionary states. Finally, neither of the intermediate SSFR groups, HCGs~7 and 22, are detected by {\em Chandra}. From the figure, we find that there are two possible evolutionary scenarios with respect to the total group SSFR: (1) the groups move from the lower right portion of Figure~\ref{fig:ssfr_lx} to the upper left, i.e., star-forming and X-ray fainter to quiescent and X-ray brighter, though perhaps not monotonically; or (2) there exists at least one more evolutionary track in which star formation in CG galaxies declines while gas is not stripped from the disks and/or heated. In particular, studies of groups similar to HCGs~7 and 22 may provide further insight.

\section{Summary}
\label{sec:sum}

We detect diffuse X-ray emission in seven of nine of the CGs in our sample with temperatures ranging from 0.6--0.72~keV and bolometric X-ray luminosities between $10^{40.4}$ and $10^{42.2}$~erg~s$^{-1}$. The groups exhibit a wide range of velocity dispersions (56--343~km~\per{s}), $\log_{10}(M_{\mathrm{H\ I}})/\log_{10}(M_{\mathrm{dyn}})$ (0.70--0.97), and morphological fractions from spiral-only groups to systems rich with E/S0 galaxies. 

Based on the hot gas morphologies, we find that the X-ray emission likely arises due to both local processes (i.e.~star formation, nuclear activity, and tidal interactions) and global processes (i.e.,~heating by the group potential). In dynamically unevolved (i.e.,~low-mass, low velocity dispersion) systems, the observable diffuse X-ray emission is dominated by local processes. The X-ray brighter groups (for a given \hi mass ratio) have emission that stems from both an extended diffuse component (i.e.,~a true intragroup medium) and galaxy-linked emission, while emission detected in the X-ray fainter groups is only associated with the individual galaxies.

HCGs~31, 42, 59, and 92 have X-ray luminosities in agreement with the predicted values from the $L_X-T$ scaling relation from clusters, though the error in the temperatures of HCGs~31 and 59 are large. Furthermore, HCGs~16 and 31 agree with the $L_X-\sigma$ relation. In both $L_X-T$ and $L_X-\sigma$, HCG~62 appears to lie within the scatter of the cluster data and is similar to fainter $L_X$ clusters. The agreement between HCGs~16 and 31 with the $L_X-\sigma$ relation are likely coincidental because the scaling relation is predicted from the Virial theorem (i.e.,~$L_X\propto\sigma^4$), and the X-ray emission from these two systems is clearly dominated by star formation rather than virilization. When the groups disagree with the cluster scaling relations, particularly in $L_X-\sigma$ space, this indicates that the groups are not simply scaled-down analogs to galaxy clusters. Furthermore, given that the hot gas in the low-mass (i.e.,~low velocity dispersion) systems is found to be isolated to the group members rather than throughout the intragroup medium, we conclude that galaxy clusters are not a proper comparison class of objects for these groups (noting the possible exception of HCG~62).

We also find that there may be a relationship between $L_X$ and how the \hi gas is distributed: preliminary evidence suggests that CGs with gas stripped from the galaxies are brighter in X-rays than groups with \hi confined to the members, possibly due to strong multi-galaxy interactions that dispersed neutral gas into the intragroup medium and triggered star formation. However, the X-ray faintest groups are also more heavily deficient in \hi implying that there may be some fraction of missing gas, possibly too cool to emit in X-rays or with too low surface brightness to detect. Finally, we note that groups dominated by local heating mechanisms have high UV+24~$\mu$m specific star formation rates, while groups with gas heated by the group potential have low SSFRs. The values of $L_X$ between these two categories spans $\sim$2 orders of magnitude and may indicate the presence of an ``X-ray gap'' in CGs similar to the SSFR and mid-IR color gaps found by \citet{tzanavaris10} and \citet{walker10,walker12}.

The faintest $L_X$ groups appear to be at very early stages in their evolution, perhaps coming together for the first time as is indicated by their low fractions of E/S0 galaxies. The influence of multi-galaxy interactions on liberating neutral gas from the galaxies and depositing it into the intragroup medium early in the group lifetime (e.g.,~as seen in HCG~16) appears to have an effect on the ability of these groups to evolve into more cluster-like systems with respect to the hot gas distribution (e.g.,~HCG~62).

\subsection{Future Work}

Expanding the sample with appropriate observations of groups from the {\em Chandra} and {\em XMM-Newton} archives is the logical next step. This will give us a larger sample with which to study the relation between the hot gas and the evolution of group environment (e.g., how the X-ray luminosity varies with \hi mass ratio). Furthermore, continuing to examine the differences between compact, loose, and fossil groups will demonstrate how efficient the compact environment processes gas. Inclusion of multiwavelength data will help to facilitate comparison of the hot gas to gas in cooler states (e.g., cold molecular). HCG~16 in particular is an interesting group worth more study; the presence of two potential M82-like superwinds in a single system presents an interesting case study for tidally induced star formation and how superwinds drive hot gas into the intragroup medium.

%In addition, we intend to examine the HCG~16 group more closely as an analog to the M81 galaxy group. The possible presence of two M82-like starbursts with potential superwinds in a single system presents an interesting case study for tidally induced star formation. We plan to examine this group in more detail with deep X-ray and optical imaging and spectroscopy to characterize the properties of the galaxies, analyze the star cluster population of the group, and search for signs of star formation and low surface brightness features within the intragroup medium (e.g.,~three sites of ongoing intergalactic star formation observed in H$\alpha$; \citealt{werk10}). With its superior spatial resolution compared to other modern X-ray observatories, deep {\em Chandra} observations will also enable us to better detect the point source population of HCG~16 for study and excise the sources from the data to characterize more robustly the build-up of hot gas in the IGM.

\acknowledgments{T.D.D. and S.C.G. thank the Natural Science and Engineering Research Council of Canada and the Ontario Early Researcher Award Program for support. This work was partially supported by the ACIS Instrument Team contract SAO~SV4-74018 (PI:~G.~Garmire). W.N.B. thanks NASA ADP grant NNX10AC99G and NSF grant AST-1108604 for support. The authors are grateful to the anonymous referee for providing thoughtful feedback that improved the manuscript. T.D.D. also thanks Tesla Jeltema, Allison R. Hill, and Alexander DeSouza for their helpful comments. The Institute for Gravitation and the Cosmos is supported by the Eberly College of Science and the Office of the Senior Vice President for Research at the Pennsylvania State University. J.C.C., I.S.K., and C.G. acknowledge funding that was provided through Chandra Award No.~GO8-9124B issued by the {\em Chandra X-ray Observatory} Center, which is operated by the Smithsonian Astrophysical Observatory under NASA contract NAS8-03060, and by the National Science Foundation under award 0908984. This research has made use of the NASA/IPAC Extragalactic Database (NED) which is operated by the Jet Propulsion Laboratory, California Institute of Technology, under contract with the National Aeronautics and Space Administration.}

\begin{appendix}
\section{Notes on Individual Groups}
\label{app:sourcenotes}

In addition to the unique aspects of the X-ray analysis and properties of the diffuse emission, we also list the center RA and Dec, shape, and dimensions of the extraction region for each group.

{\em HCG~7}.---Diffuse emission in this group was not detected above the background in the {\em Chandra} data. The circular extraction region with radius 3\farcm9 was centered at RA~=~\rascend{00}{39}{23}{9} and Dec~=~\decline{+00}{52}{15}{4}.

{\em HCG~16}.---The shallow observation of this group prevented detection of a true intragroup medium (i.e., the $L_X=2.4\times10^{40}$~erg~s$^{-1}$ IGM, corrected to our cosmology, found by \citealt{belsole03} with 45~ks of {\em XMM} data); however, hot gas associated with the individual galaxy members was detected. The extracted spectrum corresponds to the area surrounding galaxies A, B, C, and D, but does not include galaxy X, which is far removed from the group center and located far from the S3 aimpoint of the observation. The rectangular 7\farcm2$\times$3\farcm7 extraction region was centered at RA~=~\rascend{02}{09}{33}{0} and Dec~=~\decline{--10}{09}{05}{8} with PA~=~36\degree.

{\em HCG 22}.---Diffuse emission in this group was not detected above the background in the {\em Chandra} data. The rectangular 6\farcm5$\times$6\farcm9 extraction region was centered at RA~=~\rascend{03}{03}{30}{1} and Dec~=~\decline{--15}{39}{27}{3} with PA~=~22\degree. We note that this region includes a background pair of galaxies to the southeast of the main group members, however any emission from this pair is negligible and does not adversely effect the upper limit on the X-ray luminosity of HCG 22.

{\em HCG~31}.---The S/N of the detection is only marginally above the threshold required for a detection (S/N$=3$), and therefore the properties of the hot gas in this system are poorly constrained. However, this is the first detection of diffuse X-ray emission in this group. The extraction region covers the massive group members and the southern tidal tail where the majority of star formation is occurring within the group \citep{gallagher10}. Based on the reservoir of \hi gas in the group, \citet{gallagher10} predicted that conversion of 75\% of the neutral gas into stellar mass over a 150~Myr episode of star formation would generate $\sim$\e{2}{39}~erg~s$^{-1}$ of X-ray emission, however this is an order of magnitude below the observed value of $L_X$ from the {\em Chandra} data. The elliptical 2\farcm1$\times$1\farcm3 extraction region was centered at RA~=~\rascend{05}{01}{39}{7} and Dec~=~\decline{--04}{16}{15}{8} with PA~=~44\degree.

{\em HCG~42}.---We extracted spectra from both the region around galaxy~A, where the diffuse X-ray emission is most readily apparent, and from a region containing all of the massive group members. However, the count rates obtained from both extraction regions were consistent within errors. Therefore, we find that the majority of the X-ray emission in the system is associated with galaxy~A. To ensure that we do not exclude any extended emission, we compare our value with that derived from {\em ROSAT} PSPC data using the extended radius of the X-ray emission from \citet{mulchaey98}. This radius corresponds to the distance at which the X-ray emission falls to 20\% of its peak value; in HCG~42, this radius is 8\arcmin. Comparison with the flux extracted from the {\em ROSAT} data shows that our value of the flux within our extraction region matches that obtained with an 8\arcmin\ radius. The circular extraction region with radius 3\farcm4 was centered at RA~=~\rascend{10}{00}{21}{0} and Dec~=~\decline{--19}{38}{48}{5}.

{\em HCG~59}.---As for HCG~31, the X-ray emission from this group is extremely weak; however, the S/N is  sufficient to classify this group as a detection. Due to the low S/N, the values derived from model fits to the extracted spectra are poorly constrained. Though it is unlikely that the value of the temperature could span two orders of magnitude (as indicated by the upper 90\% confidence error estimate), we include this error in subsequent figures for consistency. The 3\farcm9$\times$2$\arcmin$ elliptical extraction region was centered at RA~=~\rascend{11}{48}{26}{5} and Dec~=~\decline{12}{43}{10}{2} with PA~=~0.3\degree.

{\em HCG~62}.---Previous work has found that there are cavities in the X-ray emission around HCG~62 (the result of AGN jets and lobes due to the Seyfert~2 nucleus in galaxy~A) that lack high frequency radio emission, but do have powerful low frequency emission \citep{dong10,gitti10,giacintucci11,osullivan11}. The extraction region does not include galaxy D, which is far to the south with respect to the other group members, and has negligible X-ray emission. During model fitting, we found that a model with a single MEKAL component was insufficient to properly fit the observed spectrum, therefore we model this group with two separate plasmas. We note that \citet{mulchaey98} found that the diffuse emission extends much farther from the group center than is evident in the {\em Chandra} data, and well beyond the FOV of the ACIS CCDs. Using the extended X-ray emission radius of 24\farcm2 from \citet{mulchaey98}, we found that our measured absorbed flux was a factor of 3.1 lower than that found from the {\em ROSAT} PSPC data and thus correct our measurements by this amount. Note that we apply this correction to both components of the emission in addition to the total luminosity, therefore the luminosities of the hot and cold component should be considered upper limits. In Figure~\ref{fig:kT_v_Lx}, we separately plot both the cooler and hotter components. The circular extraction region with radius 1\farcm5 was centered at RA~=~\rascend{12}{53}{06}{0} and Dec~=~\decline{--09}{12}{11}{6}.

{\em HCG~90}.---The spectral extraction region for this group is centered on galaxies~B, C, and D that are currently interacting with one another and are embedded within a halo of diffuse optical light \citep{white03}. The region excludes the brightest member (90A), which is located $\sim$6\farcm9 (68~kpc) from the other three massive, interacting group members and contains a powerful Seyfert~2 nucleus. This bright AGN complicates analysis of the diffuse emission with bright readout streaks and substantial pileup on the I3 CCD. The 5\farcm3$\times$4\farcm1 elliptical extraction region was centered at RA~=~\rascend{22}{02}{04}{5} and Dec~=~\decline{--31}{58}{51}{9} with PA~=~100\degree.

{\em HCG~92}.---The known primary source of heating for the hot gas is a shock front caused by the high velocity intruder galaxy NGC~7318B as it moves at $\sim$850~km~s$^{-1}$ through the intragroup medium \citep{pietsch97}. Numerous interactions have occurred in the group in the past $\sim$500~Myr leading to tidal tails and debris (e.g.,~\citealt{fedotov11,hwang12}). These frequent interactions in the group likely caused gas to be stripped from the member galaxies and be deposited into the intragroup medium \citep{moles97,guillard12}. We note that the presence of the shock in the intragroup medium could have non-thermal X-ray emission that is not included in the MEKAL model fit. The circular extraction region with radius 2\farcm5 was centered at RA~=~\rascend{22}{35}{59}{5} and Dec~=~\decline{33}{58}{03}{2}.
\end{appendix}

\bibliographystyle{apj}
\bibliography{ms}{}

\end{document}